\documentclass[aps,prd,twocolumn,groupedaddress,preprintnumbers]{revtex4-1}

\usepackage{amsmath,amssymb,mathtools,bm}
\usepackage{graphicx, color, hepunits}
\usepackage[dvipsnames]{xcolor}
\usepackage{float}
\usepackage{filecontents}
 \usepackage{hyperref} %Automatically links \label and \ref commands; Always load last
\hypersetup{
    colorlinks=true,       % false: boxed links; true: colored links
    linkcolor=blue,        % color of internal links
    citecolor=blue,        % color of links to bibliography
    filecolor=magenta,     % color of file links
    urlcolor=blue          % color of external links
}
\usepackage[utf8]{inputenc}
\usepackage[english]{babel}

\newcommand{\es}[2] {\begin{equation} \label{#1} \begin{split} #2 \end{split} \end{equation}}

\begin{document}
\title{Early-Universe Simulations of the Cosmological Axion}
\author{Malte Buschmann}
\author{Joshua W. Foster}
\author{Benjamin R. Safdi}
\affiliation{Leinweber Center for Theoretical Physics, Department of Physics, University of Michigan, Ann Arbor, MI 48109}
\preprint{LCTP-19-08}

\begin{abstract}
Ultracompact dark matter (DM) minihalos at masses at and below $10^{-12}$ $M_\odot$ arise in axion DM models where the Peccei-Quinn (PQ) symmetry is broken after inflation.  The minihalos arise from density perturbations that are generated from the non-trivial axion self interactions during and shortly after the collapse of the axion-string and domain-wall network.   We perform high-resolution simulations of this scenario starting at the epoch before the PQ phase transition and ending at matter-radiation equality.  We characterize the spectrum of primordial perturbations that are generated and comment on implications for efforts to detect axion DM.  We also measure the DM density at different simulated masses and argue that the correct DM density is obtained for $m_a = 25.2 \pm 11.0 \, \, \mu\mathrm{eV}$.
\end{abstract}

\maketitle

The quantum chromodynamics (QCD) axion is a well-motivated dark-matter (DM) candidate capable of producing the present-day abundance of DM while also resolving the strong \textit{CP} problem of the neutron electric dipole moment \cite{Peccei:1977ur,Peccei:1977hh,Weinberg:1977ma,Wilczek:1977pj,Preskill:1982cy,Abbott:1982af,Dine:1982ah}.  The axion is an ultralight pseudo-scalar particle whose mass primarily arises from the operator $a G \tilde G/f_a$, with $a$ the axion field, $G$ the QCD field strength, $\tilde G$ its dual, and $f_a$ the axion decay constant.  Below the QCD confinement scale, this operator generates a potential for the axion; when the axion minimizes this potential it dynamically removes the neutron electric dipole moment, thus solving the strong \textit{CP} problem. In the process the axion acquires a mass $m_a \sim \Lambda_{\rm QCD}^2 / f_a$, with $\Lambda_{\rm QCD}$ the QCD confinement scale.  The standard ultraviolet completion of the axion low-energy effective field theory is that the axion is a pseudo-Goldstone boson of a symmetry, called the Peccei-Quinn (PQ) symmetry, which is broken at the scale $f_a$~\cite{Kim:1979if,Shifman:1979if,Dine:1981rt,Zhitnitsky:1980tq,Srednicki:1985xd}.  

The cosmology of the axion depends crucially on the ordering of PQ symmetry breaking and inflation.  
If the PQ symmetry is broken before or during inflation, then inflation produces homogeneous initial conditions for axion field and generically the cosmology is relatively straightforward \cite{Marsh:2015xka}.  
In this work we focus on the more complex scenario where the PQ symmetry is broken after reheating.  
Immediately after PQ symmetry breaking, the initial axion field is uncorrelated on scales larger than the horizon, with neighboring Hubble patches coming into causal contact in the subsequent evolution of the Universe.  This leads to complicated dynamical phenomena, such as global axion strings, domain walls, and non-linear field configurations called oscillons (also referred to as axitons)~\cite{Hogan:1988mp, Kolb:1993zz,Kolb:1993hw, Kolb:1994fi,Zurek:2006sy, Enander:2017ogx, Vaquero:2018tib}.  

We perform numerical simulations to evolve the axion field from the epoch directly before PQ symmetry breaking to directly after the QCD phase transition.  Once the field has entered the linear regime after the QCD phase transition, we analytically evolve the free-field axion to matter-radiation equality.  The central motivations for this work are to (i) quantify the spectrum of small-scale ultracompact minihalos that emerges through the non-trivial axion self-interactions and initial conditions, and (ii) to determine the $m_a$ that leads to the correct DM density in this scenario.  

The post-inflation PQ symmetry breaking cosmological scenario has been the subject of considerable numerical and analytic studies.  It has been conjectured that this cosmology gives rise to ultra-dense compact DM minihalos with characteristic masses $\sim$$10^{-13}$-$10^{-11}$ $M_\odot$, though we show that the typical masses are actually smaller than this, and initial DM overdensities of order unity~\cite{Kolb:1993hw,Kolb:1993zz,Kolb:1994fi,Tinyakov:2015cgg,Davidson:2016uok,Fairbairn:2017sil,Vaquero:2018tib}.  
   In this work we compute the minihalo mass function precisely, combining state-of-the-art numerical simulations with a self-consistent cosmological picture.  Understanding this mass function is important as it affects the ways that we look for axions in this cosmological scenario.  For example, it has been claimed that microlensing by minihalos and pulsar timing surveys~\cite{Dror:2019twh} may constrain the post-inflation PQ symmetry breaking axion scenario~\cite{Fairbairn:2017sil}, but these analyses rely crucially on the form of the mass function at high overdensities and masses.
   The axion minihalos may also impact indirect efforts to detect axion DM through radio signatures~\cite{Pshirkov:2007st,Huang:2018lxq,Hook:2018iia,Safdi:2018oeu,Bai:2017feq,safdi2019xxx}. 

A precise knowledge of the $m_a$ that gives the observed DM density is of critical importance for axion direct detection experiments~\cite{Shokair:2014rna,Du:2018uak,Brubaker:2016ktl,Kenany:2016tta,Brubaker:2017rna,TheMADMAXWorkingGroup:2016hpc, Kahn:2016aff,Ouellet:2018beu,Ouellet:2019tlz,Chaudhuri:2014dla,Silva-Feaver:2016qhh}.  We find \mbox{$m_a = 25.2 \pm 11.0\, \, \mu\mathrm{eV}$}, which is within range of {\it e.g.} the HAYSTAC program~\cite{Brubaker:2017rna}.  Our axion mass estimate is similar to that found in recent simulations~\cite{Klaer:2017ond} but disagrees substantially with earlier semi-analytic estimates~\cite{Davis:1986xc,Davis:1989nj,Battye:1994au,Wantz:2009it,Hiramatsu:2010yu,Kawasaki:2014sqa,Ballesteros:2016xej}.
The minihalo mass function is also important for interpreting the results of the laboratory experiments.
If a large fraction of the energy density of DM is in compact minihalos, it is possible that the expected DM density at Earth is quite low or highly time dependent, which means that direct detection experiments would need to be more sensitive than previously thought or use an alternate observing strategy. 

The original simulations that tried to estimate the minihalo mass function were performed in~\cite{Kolb:1993hw} on a grid of size 100$^3$.
 Ref.~\cite{Kolb:1993hw} found oscillons (soliton-like oscillatory solutions) that contribute to the high-overdensity tail of the mass function.  Note that oscillons are analogous to the breather solutions found in the Sine-Gordon equation (see {\it e.g.}~\cite{Visinelli:2017ooc}).  Recently~\cite{Vaquero:2018tib} performed updated simulations on a grid of size 8192$^3$.  
  Our results expand on and differ from those presented in~\cite{Vaquero:2018tib} in many ways, such as through our initial state that begins before the PQ phase transition, measurement of the overall DM density, evolution to matter-radiation equality, and accounting of non-Gaussianities.

\begin{figure*}[tb]
\includegraphics[width=0.325\textwidth]{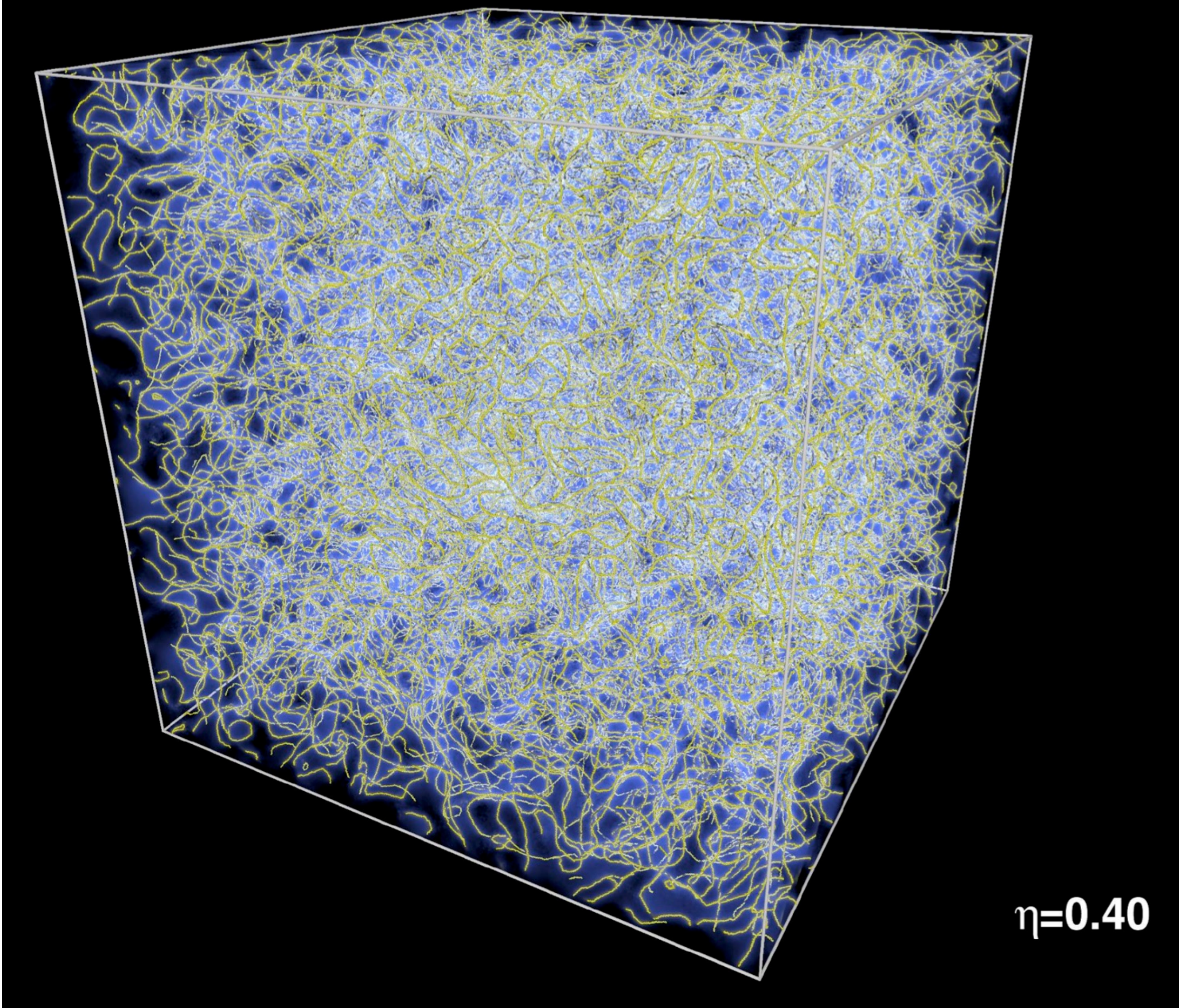}
  \includegraphics[width=0.325\textwidth]{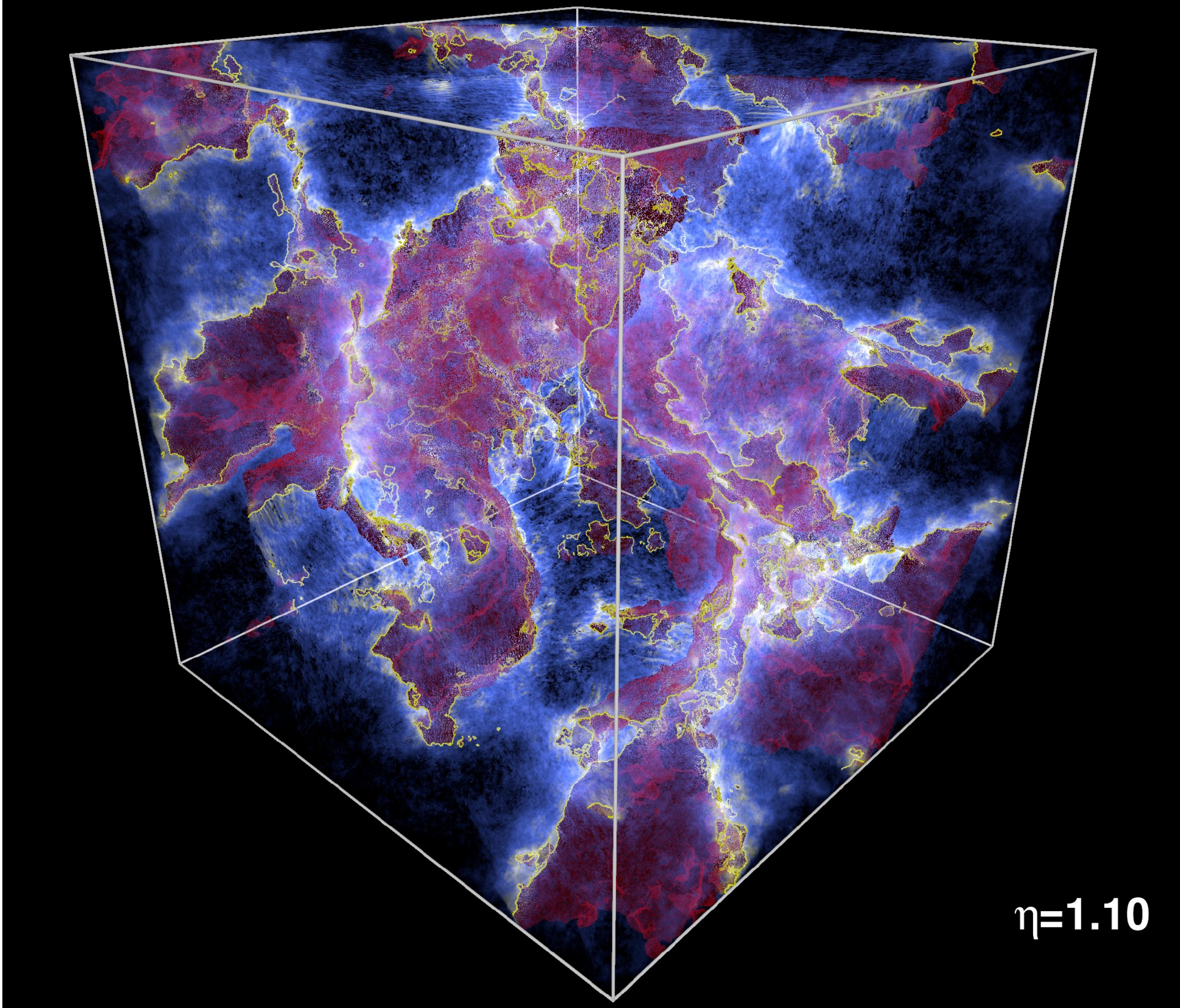}
  \includegraphics[width=0.325\textwidth]{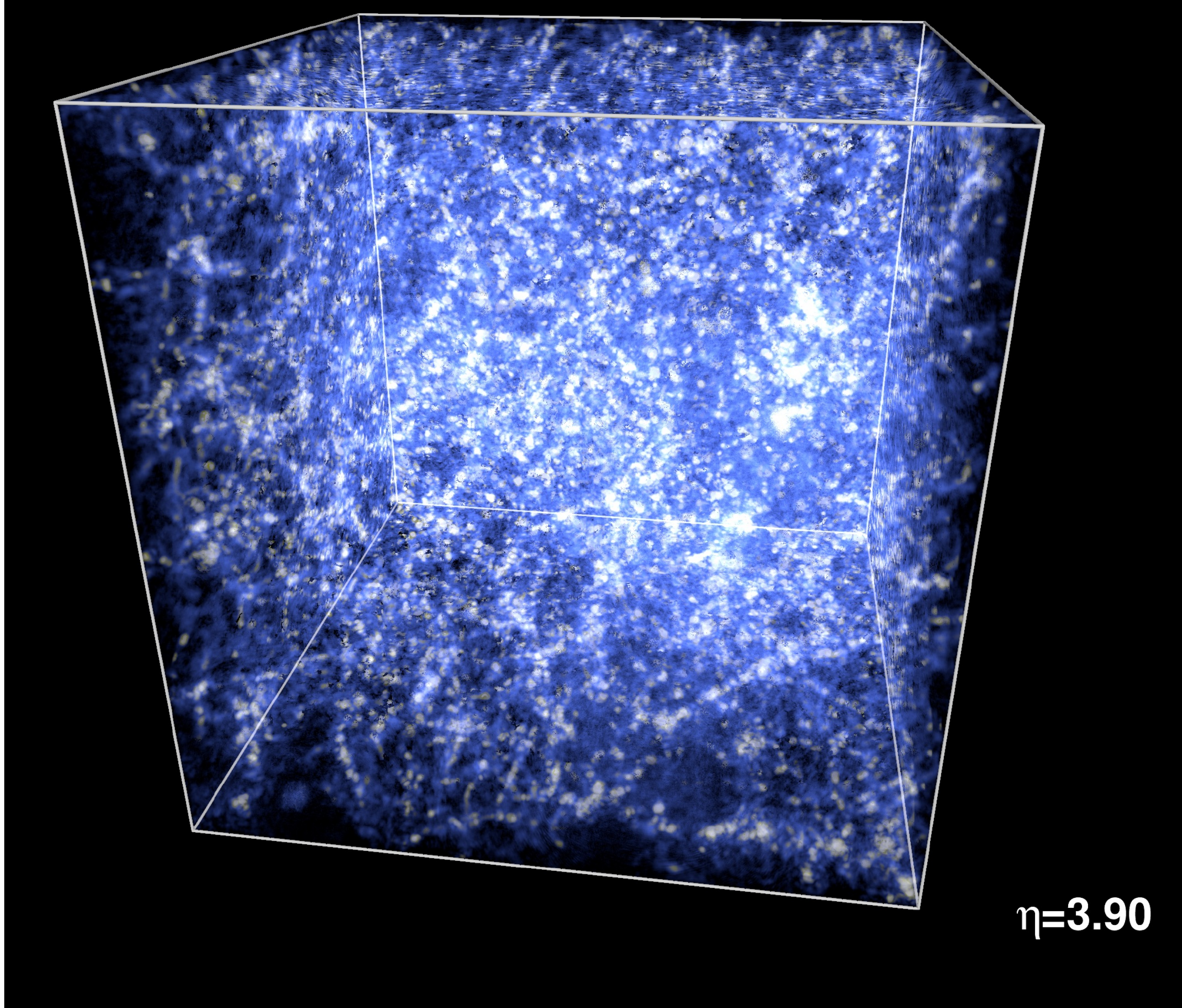}
  \vspace{-0.2cm}
  \caption{Each panel illustrates the string network (yellow strings), domain walls (red mesh), 
  and energy density of the axion field (blue-white intensity) before (left), during (middle), and after (right) the QCD phase transition (see \href{https://youtu.be/1By1DMq1EpI}{animation}).
  }
  \vspace{0.3cm}
  \label{fig:QCDevo}
\end{figure*}

\noindent
{\bf Simulation setup:} 
We begin our simulations with a complex scalar PQ field $\Phi$, with Lagrangian 
\es{eom}{
\mathcal{L}_{PQ} = &\frac{1}{2} | \partial \Phi | ^2 - \frac{\lambda}{4} \left(|\Phi|^2 - f_a^2\right)^2 - \frac{\lambda T^2}{6} |\Phi|^2 \\
&- m_a(T)^2 f_a^2 [1 - \cos \mathrm{Arg}(\Phi)],
}
with $T$ the temperature, $\lambda$ the PQ quartic coupling strength, and $m_a(T)$ the temperature-dependent axion mass generated by QCD~\cite{Hiramatsu:2012gg}.  We fix $\lambda = 1$  for definiteness, though this does not affect our final results. The parametrization of the temperature-dependent mass is adopted from the leading-order term in the fit in~\cite{Wantz:2009it}. Explicitly, the axion mass is parametrized by
\begin{equation}
m_a(T)^2 = \mathrm{min}\bigg[\frac{\alpha_a \Lambda^4}{f_a^2 (T / \Lambda)^n},  \, m_a^2\bigg],
\end{equation}
for \mbox{$\alpha_a = 1.68 \times 10^{-7}$}, \mbox{$\Lambda = 400 \, \mathrm{MeV}$} and \mbox{$n = 6.68$}, though in the Supplementary Material (SM) we consider alternate parameterizations.  The growth of the mass is truncated when it reaches its zero-temperature value, which occurs at \mbox{$T \approx 100 \, \mathrm{MeV}$} independent of the axion decay constant. The zero-temperature mass is given by \mbox{$m_a \approx 5.707 \times 10^{-5} ( 10^{11} \, \, {\rm GeV}/ f_a)$}  eV~\cite{diCortona:2015ldu}. 

For the PQ-epoch simulations we begin well before the breaking of the PQ symmetry at a time when the PQ field is described by a thermal spectrum. The simulation is performed by evolving the equations of motion on a uniformly spaced grid of side-length $L_{PQ}= 8000$ in units of $1/(a_1 H_1)$, with $a_1$ ($H_1$) the scale factor (Hubble parameter) at the temperature when $H_1 = f_a$, at a resolution of $1024^3$ grid-sites. We use a standard leap-frog algorithm in the kick-drift-kick form with an adaptive time-step size and with the numerical Laplacian calculated by the seven-point stencil.  
It is convenient to use the rescaled conformal time $\tilde \eta = \eta / \eta_1$, where $\eta_1$ is the conformal time at which point $H(\eta_1) \equiv H_1 =  f_a$.  The simulation begins at  $\tilde \eta_i = 0.0001$ and proceeds with initial time-step $\Delta \tilde \eta_i = 0.004$ until $\tilde\eta = 250$, after which a variable time-step calculated by  $\Delta\tilde\eta_i(250 / \tilde \eta)$ is used to maintain temporal resolution of the oscillating PQ fields. Convergence was tested by re-running small time intervals of the simulation at smaller time steps.
The PQ fields evolve from their initial thermal configuration until the PQ phase transition occurs at $\tilde \eta \approx 280$, after which the radial mode $|\Phi / f_a |$ acquires its vacuum expectation value (VEV). We simulate until $\tilde \eta_f = 800$ in order to proceed to a time at which fluctuations around the radial mode VEV have become highly damped.

Note that the difference in $\tilde \eta$ between $\tilde \eta = 1$ and the PQ phase transition is proportional to $\sqrt{m_{\rm pl} / f_a}$, with $m_{\rm pl}$ the Planck mass.  
The actual choice of $f_a$ here does not play an important role since we evolve the axion-string network into the scaling regime.
In the left panel of Fig.~\ref{fig:QCDevo} we show the final state of our simulation at the completion of the PQ simulation.  The string network is seen in yellow, with the blue colors indicating regions of higher than average axion density.  The length of the simulation box at this point is around $8000 /(a(\tilde \eta_f) H(\tilde \eta_f))$, and we indeed find that there is around one string per Hubble patch as would be expected in the scaling regime.

We use the final state of the PQ-epoch simulation as the initial state in our QCD-epoch simulation.  To do so we assume that the axion-string network remains in the scaling regime between the two phase transitions (see, {\it e.g.},~\cite{Hiramatsu:2010yu}). 
 Recently~\cite{Gorghetto:2018myk} found evidence for a logarithmic deviation to the scaling solution and we confirm this behavior in the SM.
  However, we perform tests to show that this deviation to scaling likely has a minimal impact on both the minihalo mass function and on the DM density, though we still assign a systematic uncertainty to our DM density estimate from the scaling violation.   

Anticipating requiring greater spatial resolution for late-times in our QCD simulation, we increased the resolution of our simulation to $2048^3$ grid-sites with a nearest-neighbor interpolation algorithm. We re-interpreted the physical dimensions of our box from side-length $L_{\rm PQ} = 8000$ in PQ spatial units to $L_{\rm QCD} = 4$ in units of $1/(a_1 H_1)$.  These units are defined such that $H_1 \equiv H( \eta_1^{\rm QCD})  = m_a(\eta_1^{\rm QCD})$ at conformal time $\eta_1^{\rm QCD}$.
  Further, we use the dimensionless parameter $\hat \eta = \eta / \eta_1^{\rm QCD}$.  While our PQ simulation ended at $\tilde \eta_f = 800$ in PQ units, the start time in the QCD phase transition is taken to be $\hat \eta_i = 0.4$ in the QCD units. Modes enter the horizon as their co-moving wavenumber becomes comparable to the co-moving horizon scale, which scales linearly with $\eta$. Therefore, by maintaining the ratio $L_{\rm PQ} /\tilde \eta_f = L_{\rm QCD} / \hat \eta_i$,  we preserve the status of our modes with respect to horizon re-entry. 
  
We then evolve the equations of motion with our initial step size now chosen to be $\Delta \hat \eta_i = 0.001$. As before, we adaptively refine our time step size, using time-step $\Delta \hat \eta_i (1.8 / \hat \eta)^{3.34}$ after $\hat \eta = 1.8$, to maintain resolution of the oscillating axion field. We simulate until $\hat \eta_f = 7.0$, periodically checking if all topological defects have collapsed. When this occurs, we switch to axion-only equations of motion for computational efficiency, since past this point the radial mode does not play an important role. 

The conformal time $\hat \eta_c$ at which the mass growth was cut off corresponds to the physical value of the axion decay constant since it relates the temperature $T_1$ at which the axion begins to oscillate and the cutoff temperature $T_c \approx 100$ MeV at which the axion reaches its zero-temperature mass. We performed simulations at five values of $\hat \eta_c$ uniformly spaced between $2.8$ and $3.6$.  These values are chosen to access different values of $f_a$ while still preserving a hierarchy between $\hat \eta_c$ and our simulation end time in order to provide sufficient time for the field to relax. At each of the five values of $\hat \eta_c$, we performed simulations at five different values of the parameter $\tilde \lambda$, defined by $\tilde \lambda \equiv \lambda f_a^2 / m_a(\hat \eta_1)^2$.
This parameter can be interpreted as the squared mass of the radial PQ mode relative to the axion mass, at conformal time $\hat \eta_1$. 
 In order for excitations of the radial mode to be well-resolved in our simulation, we require that the resolution of our simulation $\Delta \bar x$, with $\bar x = a_1 H_1 x$ and $x$ the spatial coordinate,  be such that $1 / (\hat \eta \tilde \lambda^{1/2}  \Delta \bar x) > 1$, making simulations for realistic axion parameters $\tilde \lambda$ impossible. 
We break the relation between $\tilde \lambda$ and $f_a$ and consider $\tilde\lambda=[1024,1448,3072,3584,5504]$ in order to study the impact of this parameter.
 
We illustrate three important phases of the QCD-epoch simulation in Fig.~\ref{fig:QCDevo}.  The left-most panel is the initial state discussed previously in the context of the PQ-epoch simulation final-state.  When $m_a(\hat \eta) = 3 H(\hat \eta)$ at $\hat \eta \approx 1.22$, strings grow longer and become less numerous, with domain walls forming on surfaces bounded by the strings.  This is illustrated in the middle panel, with red colors indicating domain walls. As the temperature continues to decrease with increasing $\hat \eta$, strings and domain walls tighten and decrease in size until they collapse. By $\hat \eta \gtrsim 2.0$, the network collapses in its entirety.  Shortly thereafter, we observe the formation of oscillons \cite{Kolb:1993hw, Amin:2010jq, Vaquero:2018tib}. 
We note that the oscillon field configuration is relativistic, so that near the origin of the oscillons the oscillation wavelength is $\sim$$m_a(\hat \eta)^{-1}$, which is rapidly shrinking with increasing time. After the zero-temperature mass is reached, oscillons stop shrinking and slowly dissipate at varying rates until the full field enters the linear regime.
  White regions in the right-most panel of Fig.~\ref{fig:QCDevo} denote regions of high axion energy density, which are mostly inhabited by oscillons.

 At the end of the simulation, the field has relaxed into the linear regime ({\it e.g.}, axion self-interactions are unimportant), but the field remains mildly relativistic because axion radiation is produced during the string-network collapse and during the oscillon collapse. It is therefore important to continue evolving the axion field until a time nearer to matter-radiation equality to allow the field to become non-relativistic everywhere and also to allow the compact but high-momentum overdensities to spread out. We perform this evolution analytically by exactly solving the linear axion equations of motion in Fourier space. We end this evolution shortly before matter-radiation equality ($T \sim$ keV), at which time proper velocities have frozen out but local radiation domination is preserved at all locations in our simulation box so that gravitational effects remain negligible.  

\noindent
{\bf Analysis and results.---}  
We provide \href{https://zenodo.org/record/2653964#.XPQfs9NKjOR}{Supplementary Data}~\cite{malte_buschmann_2019_2653964} containing the final state from our most realistic QCD-epoch simulation, after having performed the evolution to near matter-radiation equality.  
Note that the axion field after the QCD phase transition is highly non-Gaussian and phase-correlated at small scales and cannot accurately be reconstructed from the power spectrum.  
  In fact considering that we find large overdensities $\delta$ ($\delta \sim 10$), with $\delta = (\rho - \bar \rho) / \rho$ and $\bar \rho$ ($\rho$) the average (local) DM density, the field could not possibly be Gaussian at these scales, considering that Gaussian random fields have symmetric over and under-densities but under-densities with $\delta < -1$ would have negative DM density.    

We may try to estimate the present-day mass function by performing a clustering analysis on the final states.  In particular, we expect that the large overdensities will detach from the cosmic expansion, due to reaching locally matter-radiation equality before the rest of the Universe, and collapse onto themselves under gravity.
Thus by clustering the 3-D spatial energy density distribution from the simulation slightly before matter-radiation equality and quantifying the distribution of masses and overdensities that we find, we can make predictions for the spectrum of minihalo masses and concentrations today. 

From the final-state we construct an overdensity field $\delta(x)$, and we identify overdensities as closed regions of positive $\delta$.
 Under this
definition 50\% of the total mass is in overdensities. In practice, 
we identify these regions by first finding all positive local maxima, then recursively identifying all neighboring grid sites that are larger than 20\% of the corresponding local maxima. This threshold is necessary to have a clear boundary between different overdensities, though the final mass function is not strongly dependent on the specific choice of 20\%.   Additionally, we discard overdensities that consist of less than 80 grid sites
to avoid discretization issues in the final result. Note that we discard only about 0.8\% of the total mass 
that would otherwise be assigned to an overdensity due to the 80 grid-site limit and 20\% threshold. 
We assign to each overdensity a mass $M$ and a single mean concentration parameter $\delta$.

\begin{figure}[htb]
\includegraphics[width=0.48\textwidth]{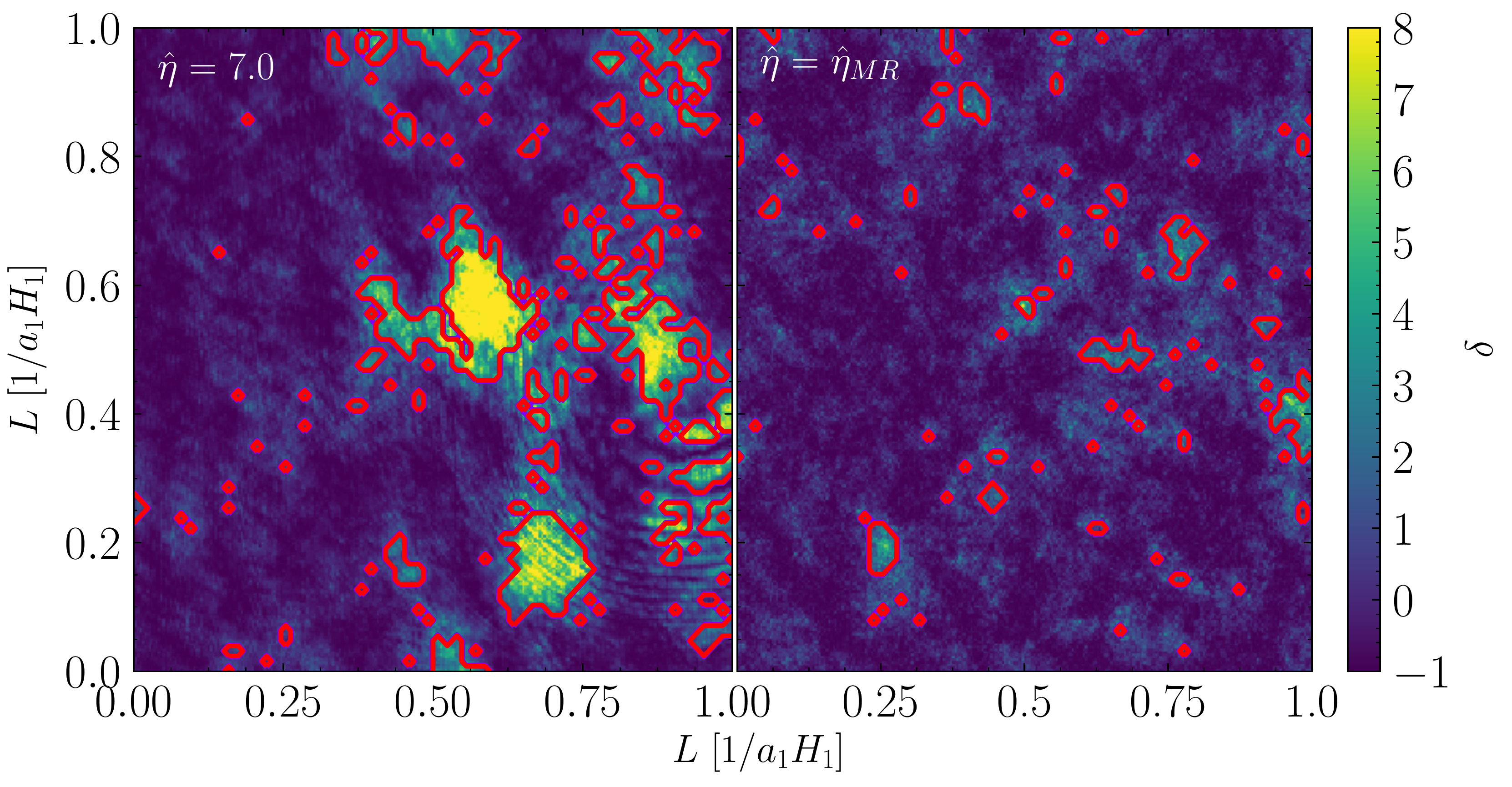}
  \caption{ (Left) A portion of a 2-D slice through the overdensity field $\delta({ x})$ at the end of the QCD stage of our most realistic simulation with $\hat\eta_c = 3.6$ and $\tilde \lambda = 5504$.  Large overdensities and rings of relativistic radiation arise from oscillon decay.  Slices through the clustered minihalos are outlined in red. (Right) As in the left panel, except the field is evolved to matter-radiation equality.  The large overdensities largely disperse and the field is everywhere non-relativistic. 
  }
  \vspace{0.3cm}
  \label{fig:cluster}
\end{figure}

An illustration of our clustering procedure is shown in Fig.~\ref{fig:cluster}.  
In that figure we show a 2-dimensional slice through the overdensity field for our most realistic simulation with $\hat\eta_c = 3.6$ and $\tilde \lambda = 5504$.  Note that in the left panel we show the field at $\hat \eta = 7$ at the end of the QCD simulation while in the right panel we show the same slice slightly before matter-radiation equality, denoted by $\hat \eta_{\rm MR} = 10^6$ and corresponding to $T \sim$ keV.   While a large overdensity left over from oscillon decay, along with corresponding rings of relativistic axion radiation, is visible in the left panel, that structure largely disperses in the subsequent evolution to $\hat \eta_{\rm MR}$.  Two-dimensional slices through the boundaries of the clustered regions are shown in red in Fig.~\ref{fig:cluster}.

\begin{figure}[htb]
\includegraphics[width=0.48\textwidth]{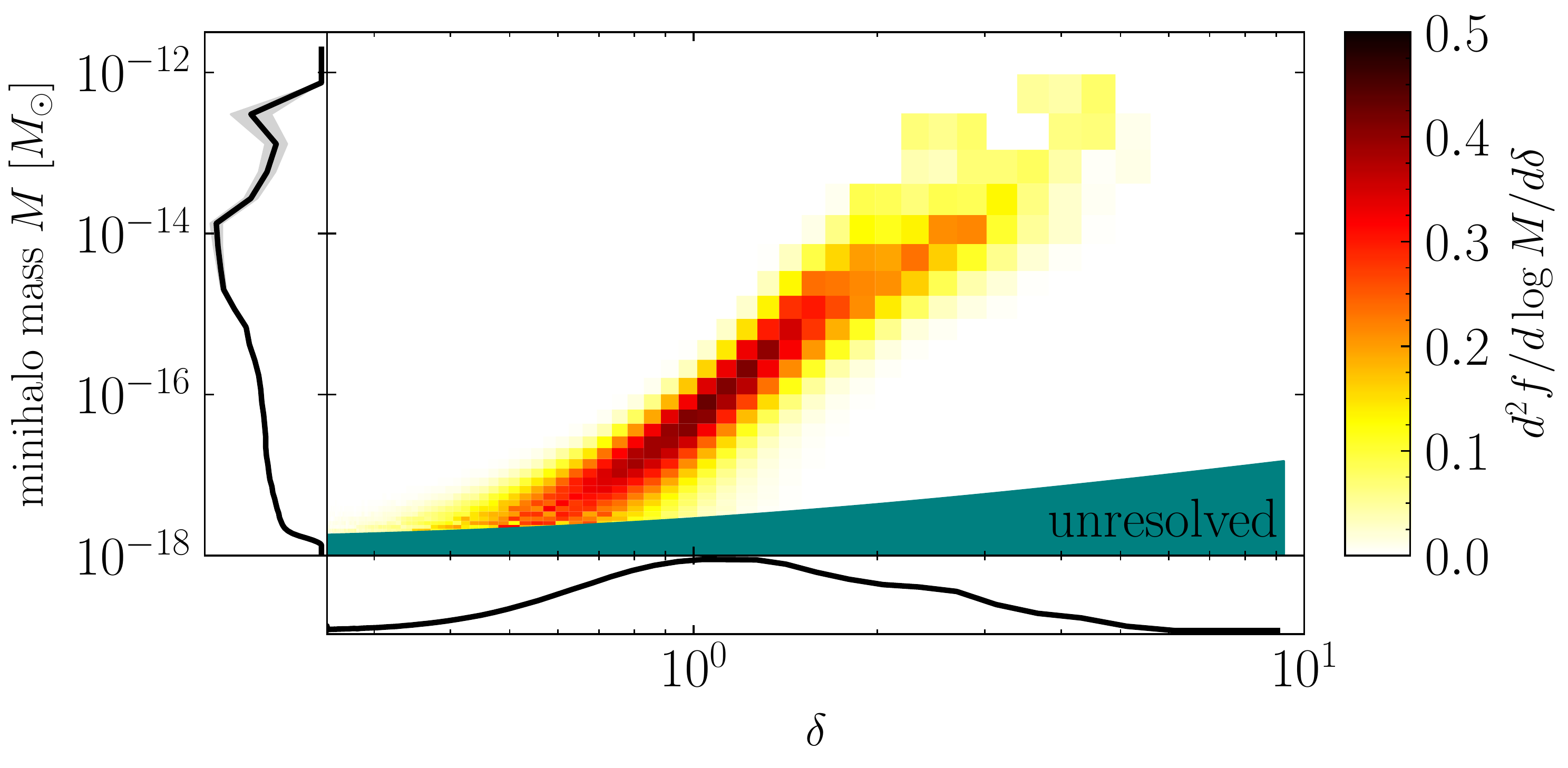} 
  \caption{Differential mass distribution for axion minihalos for our most realistic simulation, as described in Fig.~\ref{fig:cluster}, computed by clustering the overdensity field at $\hat \eta_{\rm MR}$.  The shaded ``unresolved" region denotes the parameter space that is beyond our resolution limit. Small statistical uncertainties are displayed as grey error bands.
  }
  \vspace{0.3cm}
  \label{fit:QCDaxionspectra}
\end{figure}  

We characterize the minihalo mass function through the distribution $d^2f / d(\log M) / d\delta$, where $f$ represents the fraction of mass in overdensities of mean overdensity $\delta$ and  mass $M$ with respect to the total mass in minihalos.  We compute the mass function  for all of the 25 simulations at varying $\tilde \lambda$ and $\hat \eta_c$.  To perform the extrapolation to the physical $f_a$ ($\hat \eta_c$), we use the following procedure.  First, we normalize the total DM density found in the simulation at $\hat \eta_{\rm MR}$ to the value that would give the observed DM density today.  Then we perform the clustering algorithm to determine $d^2f / d(\log M) / d\delta$.  We rescale all of the masses by $\left[ (a_1 H_1)^{\text{sim}} / (a_1 H_1)^\text{target} \right]^3$, where $(a_1 H_1)^{\text{sim}}$ is the simulated horizon size at $\hat \eta = 1$ and $(a_1 H_1)^\text{target} $ is the horizon size at the target $f_a$.  The shift accounts for the fact that the characteristic scale of the overdensities is expected to be set by the horizon volume when the axion field begins to oscillate (see, {\it e.g.},~\cite{Fairbairn:2017sil,Vaquero:2018tib} and the SM).  The effect of this shift is to move all of the masses to lower values, since the target $m_a$ is larger than those we simulate.
The resulting mass function for our most realistic simulation is shown in Fig.~\ref{fit:QCDaxionspectra}.
As we show in the SM, after applying the mass shift the mass functions appear to give relatively consistent results between the different $\hat \eta_c$, though the agreement is not perfect at high $M$.  As a result, we cannot exclude the possibility that simulating to the target $\hat \eta_c$ would give different results, especially at high masses, compared to our extrapolations.
   On the other hand, the effect of $\tilde \lambda$ appears to be minimal, since this parameter only affects the decay of the string network.

\begin{figure}[htb]
\includegraphics[width=0.48\textwidth]{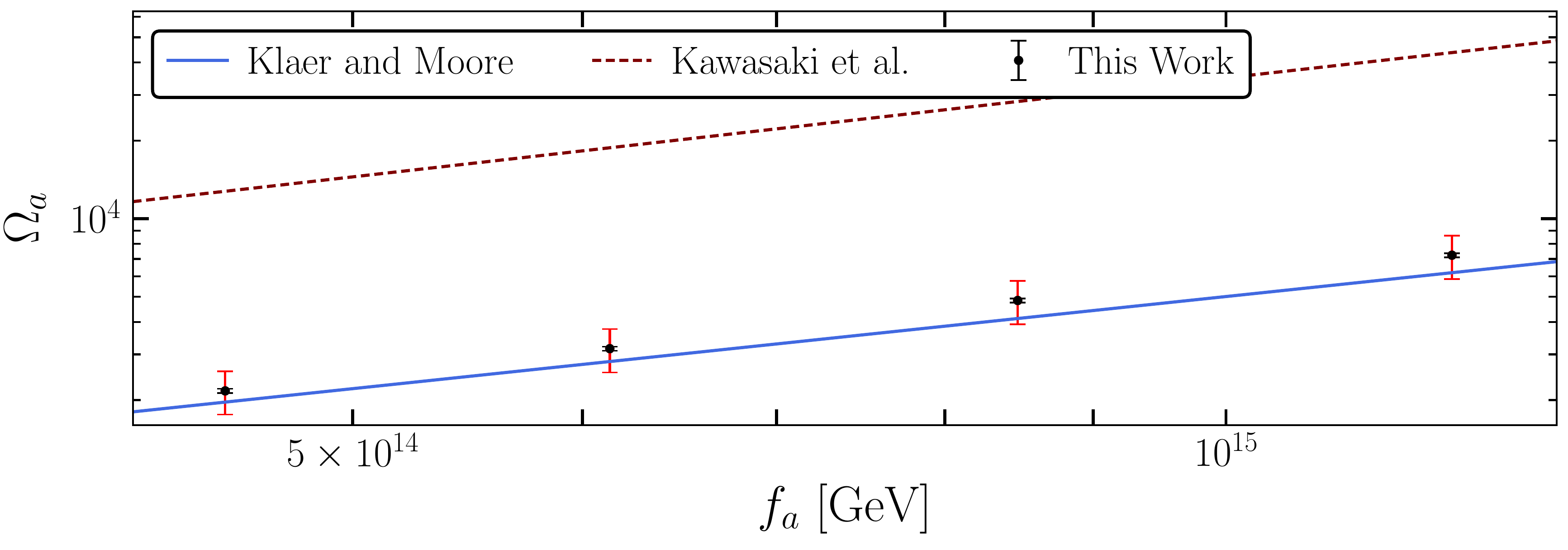} 
  \caption{The DM density $\Omega_a$ as a function of the axion decay constant $f_a$, with statistical uncertainties (black) and correlated systematic uncertainties (red) indicated, for our top four simulations.  We compare our results to those in~\cite{Klaer:2017ond} (Klaer and Moore), which agree relatively well with our own, and~\cite{Kawasaki:2014sqa} (Kawasaki et al.), which predicts significantly higher $\Omega_a$ relative to what we find.  }
  \vspace{0.3cm}
  \label{fig:DM-density}
\end{figure}

We may also compare our determinations of the total DM density produced during the QCD phase transition to previous analyses (see {\it e.g.}~\cite{Davis:1986xc,Davis:1989nj,Battye:1994au,Wantz:2009it,Hiramatsu:2010yu,Kawasaki:2014sqa,Ballesteros:2016xej,Klaer:2017ond}).  Our results are summarized in Fig.~\ref{fig:DM-density}, where we show the DM density today that we find for our top four $\hat \eta_c$, converted to $f_a$, for our most physical $\tilde \lambda$.  The uncertainties in our $\rho_a$ measurements are determined from the variance between the different $\tilde \lambda$ simulations, and while some small dependence on $\tilde \lambda$ is expected, we find that this dependence is subdominant to statistical noise and no trend is detectable in our data. We also include a conservative $10\%$ systematic uncertainty that accounts for our unphysical fixing of the effective number of degrees of freedom $g_*$ throughout our simulation, a 15\% systematic uncertainty from violations to scaling between the PQ and QCD phase transitions, and the uncertainty on the measured value of $\Omega_a$ in our Universe~\cite{Aghanim:2018eyx} (see the SM for details). 
 
In Fig.~\ref{fig:DM-density} we compare our results to the best-fit simulation result from~\cite{Klaer:2017ond}, which like us numerically evolved the axion-string system through the QCD phase transition, albeit with a different formalism, and also the semi-analytic calculations from~\cite{Kawasaki:2014sqa}.  Our results are in reasonable agreement with those in~\cite{Klaer:2017ond} and  significantly disagree with those in~\cite{Kawasaki:2014sqa}.  
 Note that we self-consistently account for all production mechanisms for axion DM in our simulation, including string decay in the few decades before the QCD phase-transition.
  It is the late-time axion production, right before the QCD phase transition, which is most important since it is the least redshifted~\cite{Kawasaki:2014sqa}.  The source of the discrepancy could be due in part to the fact that by artificially separating the production mechanisms,~\cite{Kawasaki:2014sqa} over-counted the DM density produced 
 (see~\cite{Klaer:2017ond}).  Additionally, the highly non-linear axion dynamics at the QCD epoch likely violate the number-conserving assumptions made by~\cite{Kawasaki:2014sqa}. 

    We may estimate the $f_a$ that gives the correct DM density by fitting our results to a power-law $\Omega_a \sim f_a^\alpha$.  We find the best-fit index $\alpha = 1.24 \pm 0.04$, only including statistical uncertainties, which is marginally compatible with the analytic calculations in~\cite{Kawasaki:2014sqa,Klaer:2017ond} that predict \mbox{$\alpha = (n+6)/(n+4) \approx 1.187$}.  
    Fixing $\alpha$ to the theoretical value, we find $\Omega_a = (0.102 \pm 0.02) \times (f_a/ 10^{11} \mathrm{GeV})^{1.187}$, now incorporating the correlated systematic uncertainties, which leads to the prediction that the correct DM density is achieved for $f_a = (2.27 \pm 0.33) \times 10^{11} \, \, \mathrm{GeV}$ ($m_a = 25.2 \pm 3.6 \, \, \mu\mathrm{eV}$) in agreement with~\cite{Klaer:2017ond}.  Note that if we fit for $\alpha$ instead of fixing $\alpha$ to the theoretical value we find $m_a = 17.4 \pm 4.5 \, \mu\mathrm{eV}$; the difference between the two $m_a$ estimates could be due to a systematic difference between the theoretically predicted $\alpha$ and the actual dependence of $\Omega_a$ on $f_a$.  In light of this we use the difference between the two $m_a$ estimates as an estimate of the systematic uncertainty from the extrapolation to $f_a$ below those simulated.  We additionally include a $\sim$27\% uncertainty on $m_a$ from uncertainties in the mass growth of the axion (see the SM for details), leading to the prediction $m_a = 25.2 \pm 11.0$ $\mu$eV.  
     
 \noindent
 {\bf Discussion.---}  We performed high-resolution simulations of axion DM in the cosmological scenario where the PQ symmetry is broken after inflation, starting from the epoch before the PQ phase transition and evolving the field until matter-radiation equality.  
 After matter-radiation equality one should still evolve the axion field gravitationally down to lower redshifts, which we plan to do in future work.  Our mass function is an estimate of the resulting mass function one would find after simulating the gravitational collapse.  It is possible that the true halos will be slightly larger in mass due to {\it e.g.} accretion of surrounding DM.  
 
 We may try to estimate the halo sizes based upon when we expect the halos to collapse gravitationally.  Under the assumption, for example, that the final density profile is a constant-density sphere of radius $R$ (which is likely not a good approximation but still is useful to get a sense of the halo sizes), then the halo density today was argued to be approximately $\rho \approx 140 \rho_{\rm eq} \delta^3 (\delta + 1)$, where $\rho_{\rm eq}$ is the DM density at matter-radiation equality~\cite{Kolb:1994fi}.  This implies, for example, that a $M = 10^{-14}$ $M_\odot$ subhalo with an initial average overdensity $\delta = 3$ will have a characteristic size of $\sim$$1 \times 10^6$ km.  The implications for direct and indirect axion detection efforts ({\it e.g.}, non-trivial time dependence) are likely substantial and will be the subject of future work.  One immediate implication, however, is that the axion minihalos are likely out of reach for microlensing and pulsar timing surveys~\cite{Dror:2019twh}, given the small minihalo masses.

{\it We thank Jiji Fan for collaboration on early stages of this project, and we thank Asimina Arvanitaki, Masha Baryakhtar,  Gus Evrard, Andrew Long, Nicholas Rodd, Jesse Thaler, Ken Van Tilburg, and Kathryn Zurek for useful comments and discussion. This work was supported in part by the DOE Early Career Grant de-sc0019225. JF received additional support from a Leinweber Graduate Fellowship. This research was supported in part through computational resources and services provided by Advanced Research Computing at the University of Michigan, Ann Arbor. This research used resources of the National Energy Research Scientific Computing Center (NERSC), a U.S. Department of Energy Office of Science User Facility operated under Contract No. DE-AC02-05CH11231.
}

%%%%%%%%%%%%%%%%%%%%%%%%%%%%%%%
\bibliography{Bibliography}
%\bibliographystyle{unsrt}
%%%%%%%%%%%%%%%%%%%%%%%%%%%%%%% 

\clearpage

\onecolumngrid
\begin{center}
  \textbf{\large Supplementary Material for Early-Universe Simulations of the Cosmological Axion}\\[.2cm]
  Malte Buschmann, Joshua W. Foster, and Benjamin R. Safdi\\[.1cm]
  {\itshape Leinweber Center for Theoretical Physics, Department of Physics, University of Michigan, Ann Arbor, MI 48109}
\end{center}

\onecolumngrid
%%%%%%%%%% Merge with supplemental materials %%%%%%%%%%
\setcounter{equation}{0}
\setcounter{figure}{0}
\setcounter{table}{0}
\setcounter{section}{0}
\setcounter{page}{1}
\makeatletter
\renewcommand{\theequation}{S\arabic{equation}}
\renewcommand{\thefigure}{S\arabic{figure}}

This Supplementary Material contains additional results and explanations of our methods that clarify and support the results presented in the main Letter. We begin with a detailed explanation of the equations of motion and initial conditions used in our simulations. Next, we present extended results for the overdensity spectrum and DM density.  We then present a modified simulation that allows us to quantify the systematic uncertainty in the DM density determination by assuming a fixed number of relativistic degrees of freedom.  Additionally, we quantify the uncertainty on the DM density induced by uncertainties in the mass growth of the axion, and finally we consider the effects of violations to the scaling solution on our final results.  

\section{Simulation Equations of Motion}\label{sec:SupEOM}
Our phenomenological Lagrangian describing the PQ field is adopted from the construction of \cite{Hiramatsu:2012gg} and is of the form 
\begin{equation}
\mathcal{L}_{PQ} = \frac{1}{2} | \partial \Phi | ^2 - \frac{\lambda}{4} \left(|\Phi|^2 - f_a^2\right)^2 - \frac{\lambda T^2}{6} |\Phi|^2 
- m_a(T)^2 f_a^2 [1 - \cos \mathrm{Arg}(\Phi)],
\end{equation}
where $\Phi$ is the complex PQ scalar, $T$ is the temperature, $\lambda$ is the PQ quartic coupling strength, $f_a$ is the PQ-scale identified as the axion decay constant, and $m_a(T)$ is the temperature-dependent axion mass \cite{Hiramatsu:2012gg}. The parametrization of the temperature-dependent mass is adopted from the leading order term in the fit in \cite{Wantz:2009it}. Explicitly, the axion mass is parametrized by
\begin{equation}
m_a(T)^2 = \mathrm{min}\bigg[\frac{\alpha_a \Lambda^4}{f_a^2 (T / \Lambda)^n},  \, m_a\bigg],
\end{equation}
for $\alpha = 1.68 \times 10^{-7}$, $\Lambda = 400 \, \mathrm{MeV}$ and $n = 6.68$. The growth of the mass is truncated at $T \approx 100 \, \mathrm{MeV}$. The zero-temperature mass is given by
\begin{equation}
m_a^2 = \frac{m_\pi^2 f_\pi^2}{f_a^2} \frac{m_u m_d}{(m_u + m_d)^2},
\end{equation} 
where $m_\pi$ is the pion mass, $f_\pi$ is the pion decay constant, $m_{u/d}$ is the up/down quark mass. Details of the temperature-dependent axion mass, or equivalently, the topological susceptibility, remain uncertain, especially at low temperatures.  Note that we do not explore here how our results are affected by uncertainties in the temperature-dependent axion mass, though doing so is a worthwhile direction for future work.

Decomposing the complex scalar as $\Phi = \phi_1 + i \phi_2$, and assuming a radiation-dominated cosmological background, leads to equations of motion in metric coordinates of the form 
\begin{gather}
\ddot \phi_1+ 3 H \dot \phi_1 - \frac{1}{R^2} \nabla^2 \phi_1 + \frac{1}{3} \lambda  \phi _1 \bigg[3 \left(\phi _1^2+\phi _2^2 - f_a^2\right)+T^2\bigg] -\frac{m_a(T)^2 \phi _2^2 }{\left(\phi _1^2+\phi _2^2\right){}^{3/2}} = 0 \\ 
\ddot \phi_2 + 3 H \dot \phi_2 - \frac{1}{R^2} \nabla^2 \phi_2  + \frac{1}{3} \lambda  \phi _2 \bigg[3 \left(\phi _1^2+\phi _2^2 - f_a^2\right)+T^2\bigg] + \frac{m_a(T)^2 \phi _1 \phi _2}{\left(\phi _1^2+\phi _2^2\right)^{3/2}}= 0 \,.
\label{eq: full}
\end{gather}
Over temperatures $T \gtrsim 100 \, \mathrm{MeV}$, the number of relativistic degrees of freedom $g_*$ in the Standard Model is expected to vary only mildly. For simplicity, we therefore assume $g_* = 81$, which is a typical value adopted at high temperatures (though later in the SM we explore the systematic uncertainty introduced by this assumption). 
It is useful to define a dimensionless conformal time $\hat \eta$ such that 
\begin{equation}
\hat \eta = \frac{R}{R(T = T_1)} = \frac{R}{R_1} = \left(\frac{t}{t_1}\right)^{1/2} \,,
\label{eq:etaref}
\end{equation}
where $R$ is the scale factor and the time $t_1$ (with $T(t_1) \equiv T_1$) is a reference time that will be defined differently in the PQ and QCD epoch simulations. 

The axion-mass term is not included in our PQ-epoch simulations.  In our QCD-epoch simulations, on the other hand, the mass term is included and drives the dynamics.  In this case, the mass grows until the cutoff temperature $T_c$ at which point the axion mass reaches its zero-temperature value; the corresponding conformal time is given by $\hat \eta_c = R(T = T_c) / R_1$.  Rewriting~\eqref{eq: full} with the dimensionless coordinates, we then find
\begin{gather}
\psi_1'' + \frac{2}{\hat \eta} \psi_1' - \bar \nabla^2 \psi_1 + \frac{1}{H_1^2} \left[ \lambda   \psi _1 \bigg(\hat \eta^2 f_a^2 \left(\psi _1^2+\psi _2^2 - 1\right)+ \frac{1}{3} T_1^2\bigg) -m_a^2(T_1) \hat \eta^2  \mathrm{min}(\hat \eta, \hat \eta_{c})^{n}\left(\frac{ \psi _2^2 }{\left(\psi _1^2+\psi _2^2\right){}^{3/2}} \right)\right]= 0 \\
\psi_2'' + \frac{2}{\hat \eta}\psi_2' - \bar \nabla^2 \psi_2 + \frac{1}{H_1^2} \left[\lambda \psi _2 \bigg(\hat \eta^2 f_a^2 \left(\psi _1^2+\psi _2^2 - 1\right)+ \frac{1}{3} T_1^2\bigg)+  m_a^2(T_1)\hat \eta^2 \mathrm{min}(\hat \eta, \hat \eta_{c})^{n}\left( \frac{ \psi _1 \psi _2}{\left(\psi _1^2+\psi _2^2\right)^{3/2}} \right)\right]= 0 \,,
\end{gather}
where $\phi = f_a \psi$, primes denote derivatives with respect to $\hat \eta$, and the spatial gradient is taken with respect to $\bar x = a_1 H_1 x$. 

\subsection{The PQ Epoch}

Simulations in the PQ epoch occur at $T \gg \Lambda_{\rm QCD}$ and so the temperature-dependent axion mass may be neglected. We therefore take our equations of motion to be
\begin{gather}
\psi_1'' + \frac{2}{\tilde\eta} \psi_1' - \bar \nabla^2 \psi_1 +   \lambda   \psi _1\left[\tilde\eta^2 \left(\psi _1^2+\psi _2^2 - 1\right)+ \frac{T_1^2}{3 f_a^2}\right] = 0 \\
\psi_2'' + \frac{2}{\tilde\eta}\psi_2' - \bar \nabla^2 \psi_2 +\lambda \psi _2  \left[\tilde\eta^2 \left(\psi _1^2+\psi _2^2 - 1\right)+ \frac{T_1^2}{3 f_a^2}  \right]= 0 \,,
\end{gather}
and we fix $\tilde\eta = 1$ to be the time at which $H_1 = f_a$.  Note that for our PQ-epoch simulations we refer to $\hat \eta$, defined in~\eqref{eq:etaref}, as $\tilde \eta$ in order to avoid confusion with the dimensionless conformal time $\hat \eta$ used in the QCD-epoch simulations.  The ratio $(T_1 / f_a)^2$ is determined by
\begin{equation}
\left(\frac{T_1}{f_a}\right)^2 \approx 8.4 \times 10^5 \left(\frac{10^{12} \, \mathrm{GeV}}{f_a}\right) \,.
\end{equation}
In principle, it would seem that axions of different decay constants would require different simulations in the PQ epoch. However, this ratio is degenerate with our choice of physical box size and dynamical range in $\tilde\eta$ in a particular simulation, allowing us to perform only one PQ simulation and interpret its output as the initial state of the axion field for several different values of $f_a$.  The key assumption behind this, however, is that at late times after the PQ phase transition the field enters the scaling regime so that we may reinterpret the output of the PQ simulation in the appropriately rescaled box as the initial state of the QCD simulation at much lower temperatures. 
Note that the value of $\lambda$ is a free parameter, which we naturally choose to be $\lambda = 1$ though it has little effect.

\subsubsection{Initial Conditions for a PQ Scalar} 
\label{sec:PQinitial}
We generate initial conditions for our PQ scalar by taking it to be described by a thermal distribution characterized by the temperature $T$ at the initial early time.  As can be read off from the Lagrangian, each of the two fields has an effective mass of the form
\begin{equation}
m_{\rm eff}^2 = \lambda \left( \frac{T^2}{3} - f_a^2 \right) \,.
\end{equation}
Correlation functions of the initially-free massive scalar fields are given by
\begin{align}
\langle \phi_i(x) \phi_j(y) \rangle &= \delta_{ij} \int \frac{dk}{2 \pi}\frac{n_k}{\omega_k} e^{i k \cdot(x - y)} \\
\langle \dot \phi_i(x) \dot \phi_j(y) \rangle &= \delta_{ij} \int \frac{dk}{2 \pi}n_k \omega_k e^{i k \cdot(x - y)} \\
\langle \dot \phi_i(x)  \phi_j(y) \rangle &= 0 \,,
\end{align}
where overdots denote differentiation with respect to time, and we have defined
\begin{equation}
n_k = \frac{1}{e^{\omega_k / T}-1}, \qquad \omega_k = \sqrt{k^2 + m_{\rm eff}^2}.
\end{equation}
In momentum space, these correlation functions take the form
\begin{align}
\langle \phi_i (k) \phi_j(k') \rangle &= \frac{2 \pi n_k}{\omega_k}\delta(k+k') \delta_{ij}\\
\langle \dot \phi_i (k) \dot \phi_j(k') \rangle &=  2 \pi n_k\omega_k\delta(k+k') \delta_{ij}\\
\langle \dot \phi_i(k) \phi_j(k') \rangle &= 0  \,.
\end{align}
Our simulations occur on a discrete lattice of finite size, so the correlation functions above lead to initial conditions set by a realization of a Gaussian random field specified in Fourier space by
\begin{align}
\qquad \langle \phi_i(k)  \rangle &= 0, \qquad \langle | \phi_i(k) |^2 \rangle = \frac{n_k}{\omega_k}L,  \\
\langle \dot \phi_i(k) \rangle &= 0, \qquad\langle | \dot \phi_i(k) |^2 \rangle = n_k \omega_k L \,.
\end{align}
Note that we include the 50 lowest $k$-modes in each of the three directions when constructing the initial conditions, and we have verified that including more modes does not affect our results.

\subsection{Early Times in the QCD Epoch}

During the QCD epoch, $T \sim \Lambda_{\rm QCD}$, and so the axion mass is non-negligible. Here, we define $\hat\eta = 1$ to be the time at which $H_1 = m_a(T_1)$, with the axion field beginning to oscillate shortly thereafter when $m_a = 3 H$.  The equations of motion are then given by
\begin{gather}
\psi_1'' + \frac{2}{\hat\eta} \psi_1' - \bar \nabla^2 \psi_1 + \tilde \lambda \hat\eta^2 \psi_1 (\psi_1^2 + \psi_2^2 - 1)- \mathrm{min}(\hat\eta, \hat\eta_{c})^{n}\hat\eta^2 \left( \frac{\psi_2^2}{(\psi_1^2 + \psi_2^2)^{3/2}}\right)= 0 \\
\psi_2'' + \frac{2}{\hat\eta} \psi_2' - \bar \nabla^2 \psi_2 + \tilde \lambda \hat\eta^2 \psi_2 (\psi_1^2 + \psi_2^2 - 1)+ \mathrm{min}(\hat\eta, \hat\eta_{c})^{n}\hat\eta^2 \left( \frac{\psi_1\psi_2}{(\psi_1^2 + \psi_2^2)^{3/2}}\right)= 0 \,,
\end{gather}
where we have neglected the $T_1$ contribution to the PQ scalar mass as it is small compared to $f_a$. The parameter $\tilde \lambda$ is defined by 
\begin{equation}
\tilde \lambda = \lambda\left(\frac{ f_a}{m_a(T_1)}\right)^2
\end{equation}
and can be interpreted as the squared mass of the radial mode $|\Phi / f_a |$.  For physical parameters we expect $\tilde \lambda \gg 1$, though in practice we find that the final results are relatively independent of $\tilde \lambda$ for moderately sized values of the parameter, as described in the main text and later in the SM.  Indeed, our choices for $\tilde \lambda$ allow us to resolve the radial mode mass by more than a few grid-spacings, satisfying the requirement of~\cite{Gorghetto:2018myk} to accurately study the axion spectrum from string radiation, unlike~\cite{Vaquero:2018tib}. There exist additional criteria on the largeness of $\tilde \lambda$ such that the metastability of topological defects is preserved despite the unphysical smallness of simulated $\tilde \lambda$ in comparison with the rapidly increasing axion mass. At all times prior to expected defect collapse, our choices of $\tilde \lambda$ satisfy the simplest construction of these conditions~\cite{Vaquero:2018tib}, with our choice of $\tilde \lambda = 5504$ satisfying the most stringent criteria established in~\cite{Fleury:2015aca}. We note that we are largely unable to differentiate between simulations at any two particular values of $\tilde \lambda$, and that our choice of values appear to have minimal impact, as illustrated further below.

\subsection{Late Times in the  QCD Epoch}
The presence of topological defects in the axion field at early times during the QCD epoch requires that we fully simulate both degrees of freedom of the PQ field. Once the topological defects have collapsed, however, we are free to use the axion-only equations of motion. Our axion is defined by $a = f_a \mathrm{arctan2}(\phi_1, \phi_2)$ and has the Lagrangian
\begin{equation}
\mathcal{L} = \frac{1}{2} (\partial a)^2 - m_a^2(T) f_a^2 \left[ 1- \cos\left(a \over f_a \right) \right]\,,
\end{equation}
along with corresponding equations of motion
\begin{equation}
\theta'' + \frac{2}{\hat\eta} \theta' - \bar \nabla^2 \theta + \mathrm{min}(\hat\eta, \hat\eta_{c})^{n}\hat\eta^2 \sin\theta = 0 \,.
\end{equation}
Above, we define $\theta = a / f_a$. Evolving these equations of motion is formally equivalent to freezing out excitations of the radial mode by taking $\tilde \lambda \rightarrow \infty$, which more accurately recovers the true physics of the evolution of the axion field for realistic values of $f_a$. Note that the coordinate $\bar x$ and $\hat\eta$ here are identical to those used in evolving the two degrees of freedom of the complex scalar performed prior to defect collapse.

\subsection{Analytically Evolving in the Fixed-Mass Small-Field Limit}
At late times when the axion mass has reached its zero-temperature value and the axion field has redshifted considerably so that $|\theta| \ll 1$, the equations of motion are linear and well-approximated by 
\begin{equation}
\theta'' + \frac{2}{\hat\eta} \theta' - \nabla^2 \theta + \hat\eta_c^n \hat\eta^2 \theta = 0.
\end{equation}
We may solve this equation analytically by going to Fourier space and adopting an ansatz for the solution as 
\begin{equation}
\theta(\hat\eta) = f(\hat\eta) \exp(i \mathbf{k} \cdot \mathbf{x}) \,.
\end{equation}
This ansatz leads to the equation
\begin{equation}
f''(\hat\eta )+\frac{2 f'(\hat\eta)}{\hat\eta }+ f(\hat\eta ) \left(\hat\eta ^2 \hat\eta _c^n+\mathbf{k}^2\right)=0 \,,
\end{equation}
which has the general solution
\begin{equation}
f(\hat\eta) = \frac{\exp(-\frac{i}{2}\hat\eta ^2 \hat\eta _c^{n/2}) }{\hat\eta} \left[ C_1 H_{-\frac{1}{2} \hat\eta _c^{-n/2}
   \left(\hat\eta _c^{n/2}+i \mathbf{k}^2\right)}\left(\sqrt[4]{-1} \hat\eta  \hat\eta
   _c^{n/4}\right)+C_2 \,_1F_1\left(\frac{1}{4} \hat\eta _c^{-n/2} \left(\hat\eta _c^{n/2}+i\mathbf{k}^2\right);\frac{1}{2};i \hat\eta ^2 \hat\eta _c^{n/2}\right) \right] \,,
\end{equation}
for coefficients $C_1$ and $C_2$ determined by boundary conditions, and where $H_n$ and $\,_1F_1$ are the analytic continuations of the Hermite polynomials and the confluent hypergeometric function of the first kind, respectively. From this analytic solution, we can transfer late-time field configurations from our simulation to arbitrary large $\hat\eta$. Differentiation with respect to $\hat\eta$ may be straightforwardly performed to find $f'(\hat\eta)$ at large $\hat\eta$ as well. The computation of the analytically continued Hermite polynomials and hypergeometric functions was performed with the python package \texttt{mpmath}.

\begin{figure*}[htb]
\includegraphics[width=.49\textwidth]{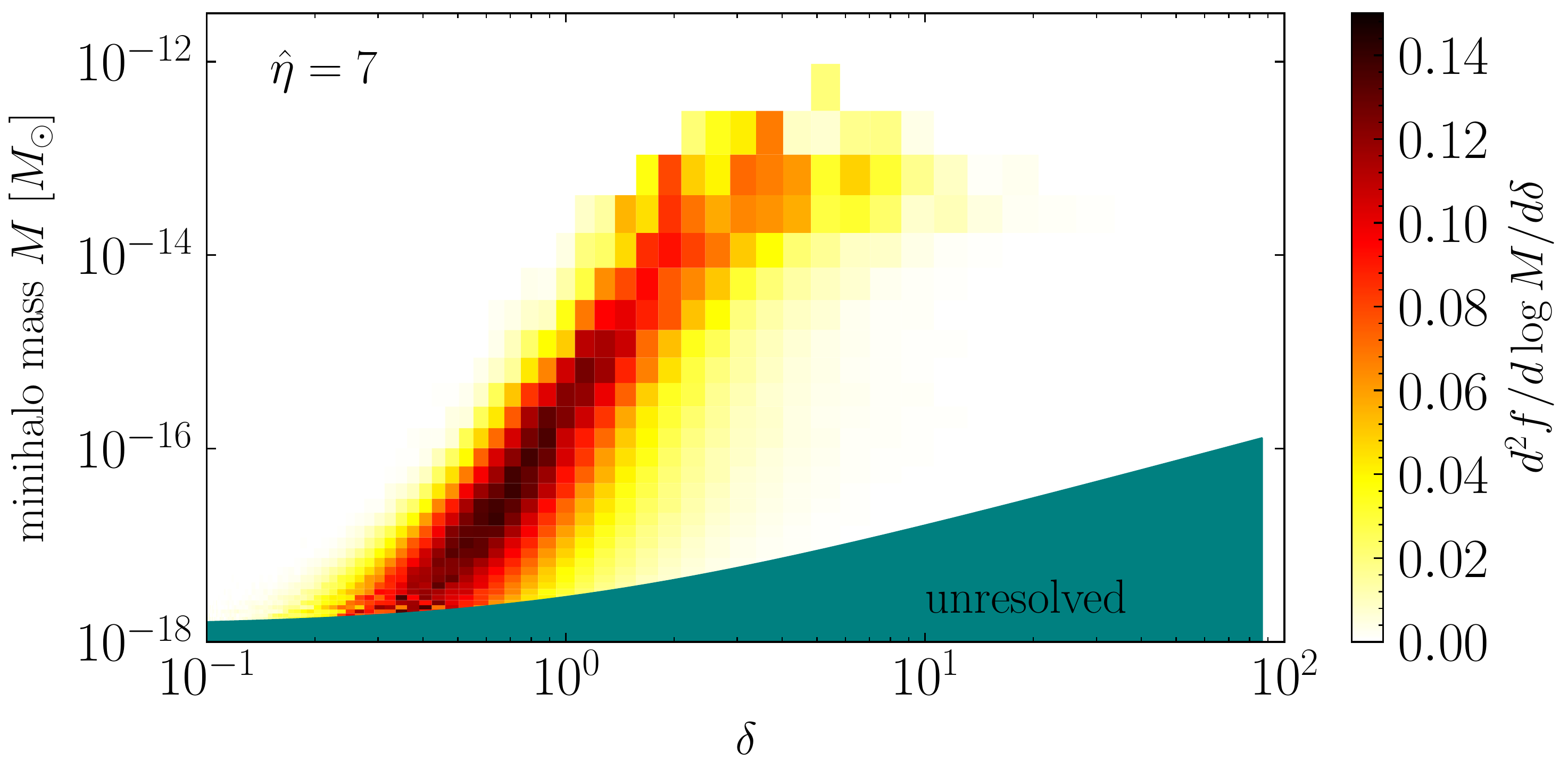}
\includegraphics[width=.49\textwidth]{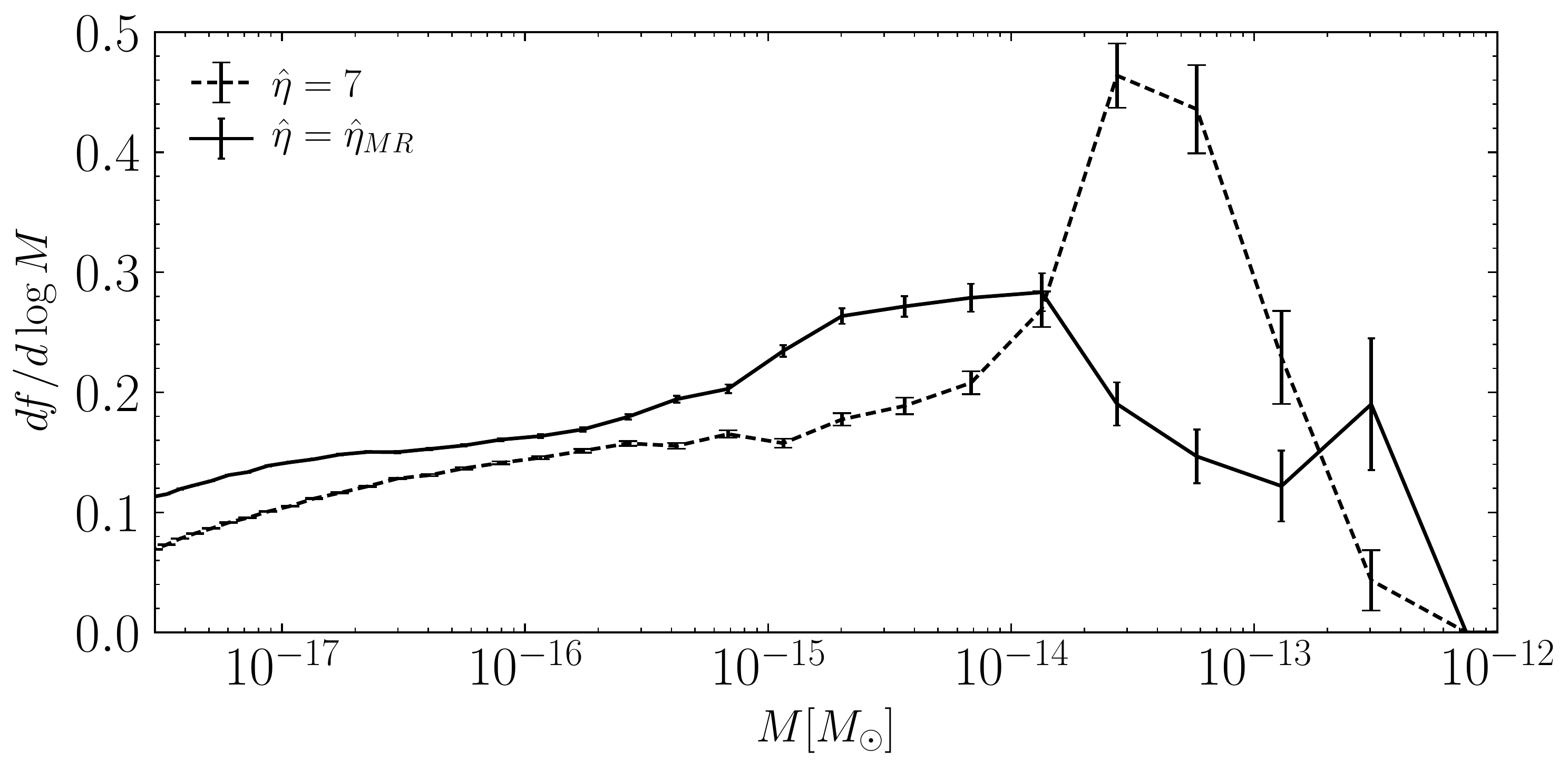}\\
\includegraphics[width=.49\textwidth]{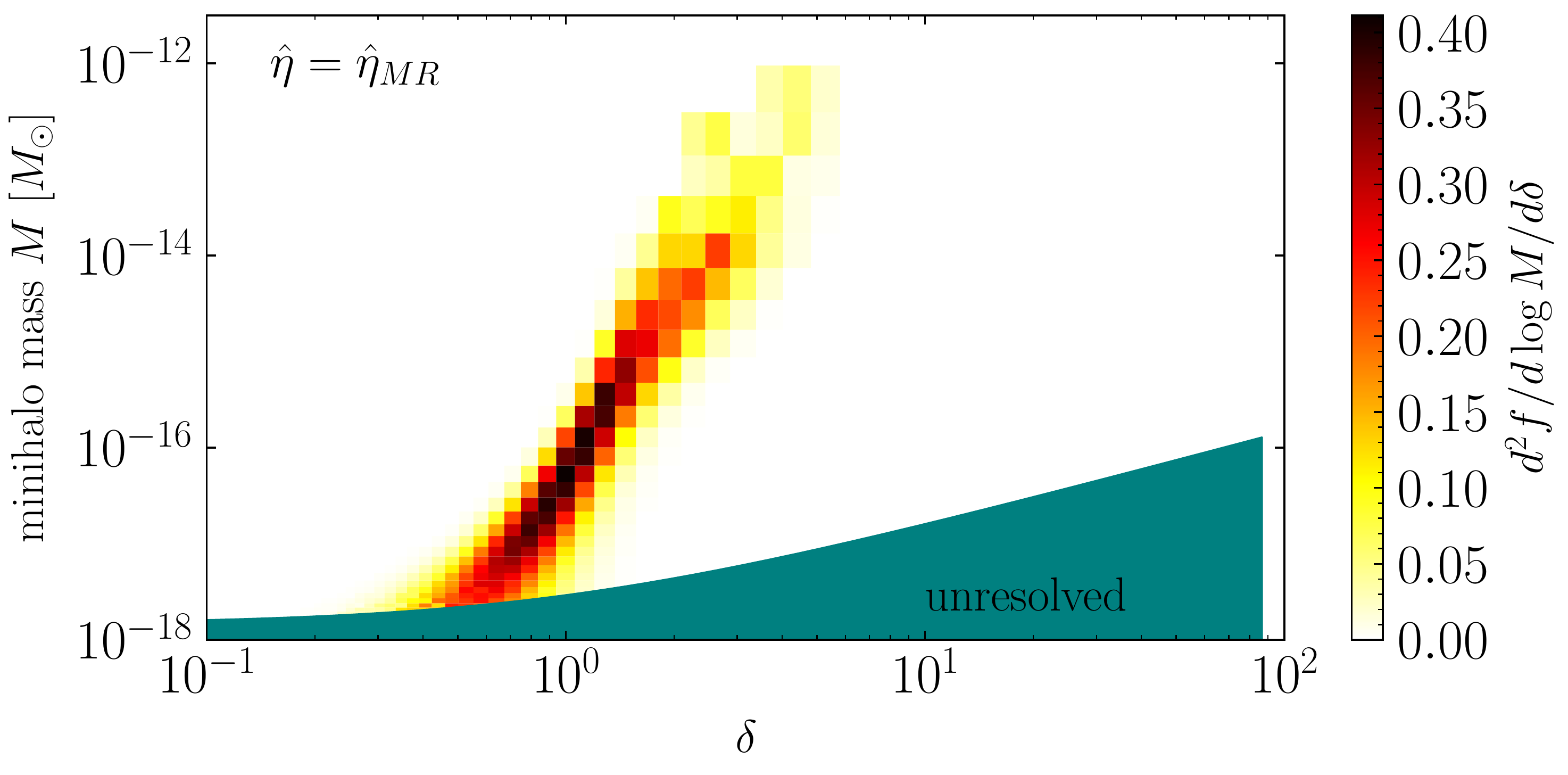}
\includegraphics[width=.49\textwidth]{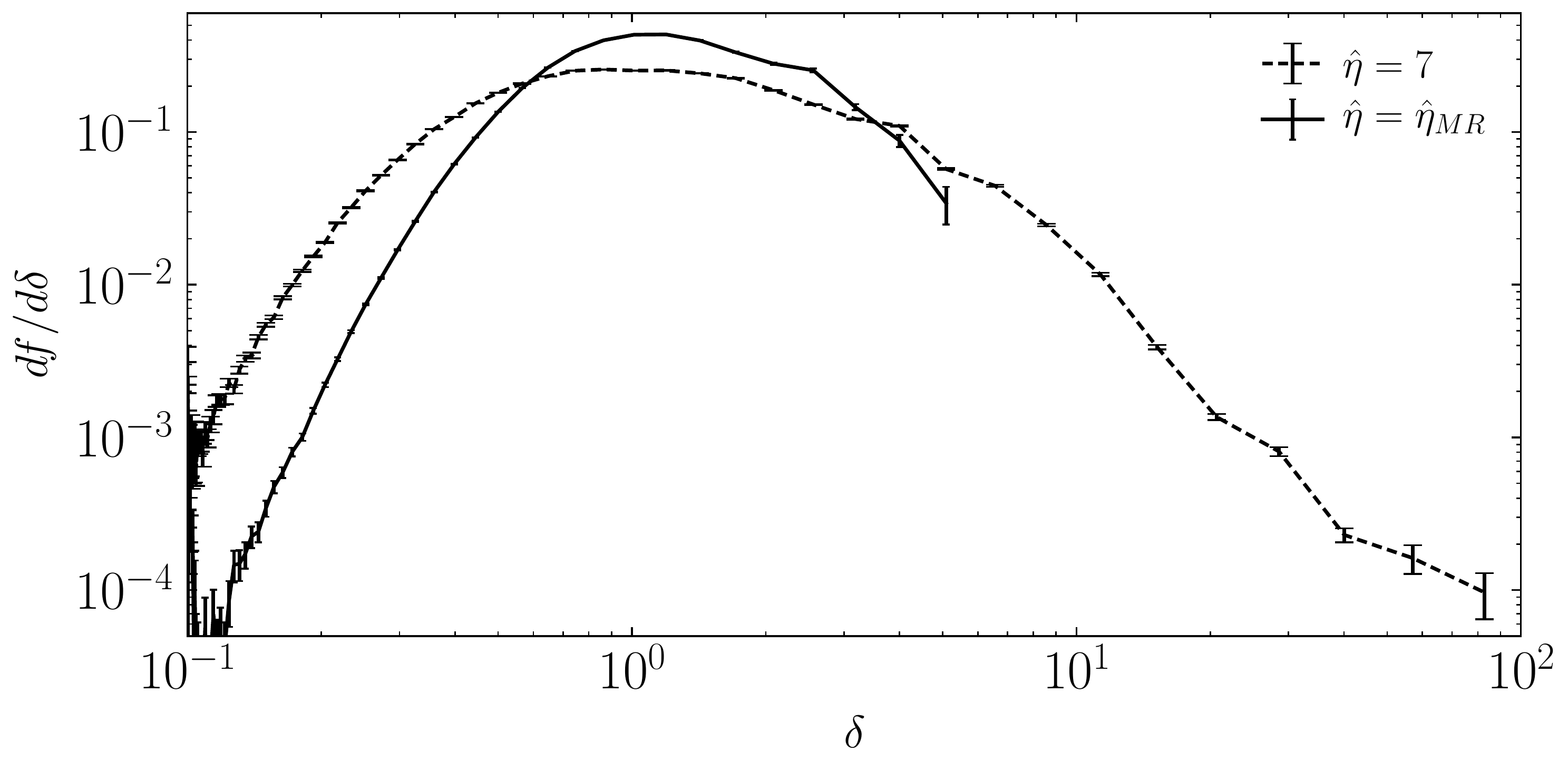}
  \caption{Double differential mass fractions for axion minihalos as a function of the concentration parameter $\delta$ and mass $M$.  In the top left we compute that mass function using the field immediately after the QCD phase transition, at $\hat \eta = 7$, while in the bottom left we use the more correct procedure of first evolving to $\hat \eta = \hat \eta_{\rm MR}$ before performing the clustering procedure.  Evolving to matter-radiation equality gives the most over-dense regions time to expand and results in less dense overdensities, as compared to the incorrect procedure shown in the top left.  This is perhaps even more apparent in the single differential mass fractions as a function of the mass $M$ (top right) and concentration parameter $\delta$ (bottom right). These results are based on our most realistic simulation with $\hat\eta_c=3.6$ and $\tilde\lambda=5504$. Error bars are statistical, and we do not extend the $df/d\log M$ curves to lower masses as we are unable to resolve those properly.}
  \vspace{0.3cm}
  \label{fig:MDelta7}
\end{figure*}

We directly compare the differential mass spectrum at $\hat\eta=7$ with the same field analytically evolved  to $\hat\eta=\eta_\text{MR}$ in Fig.~\ref{fig:MDelta7}. While the basic differential shape is the same, the $\hat\eta=7$ results have a much wider distribution in $\delta$. In particular, all overdensities above $\delta>10$ have vanished by the time matter-radiation equality is reached. However, the peak of the distribution is still around $\delta=1$.  Evolving the fields down to matter-radiation equality is important because many of the modes are generated with high momentum at the QCD epoch, causing the large overdensities to disperse by the time of matter-radiation equality.  

\section{Studying the (Over)Density Field}
Our interest in this work is studying the energy density field $\rho$ and the overdensity field $\delta = (\rho - \bar \rho) / \rho$ realized in the axion field from our simulations.  The axion energy density for the axion field $a = f_a \theta$ is computed by the Hamiltonian density 
\begin{equation}
\mathcal{H} = f_a^2 \left[\frac{1}{2}\dot \theta^2 + \frac{1}{2 R^2}( \nabla \theta)^2 + m_a^2(1-\cos\theta)\right], 
\end{equation}
which can be rewritten in simulation units as
\begin{equation}
\mathcal{H} = m_a^2 f_a^2\left[ \frac{\theta'^2 + (\bar\nabla \theta)^2}{2 \hat\eta_c^{6.68} \hat\eta^2} + (1 - \cos \theta) \right] \,,
\end{equation} 
assuming $\hat\eta > \hat\eta_c$. At late times, the Hamiltonian is approximately  
\begin{equation}\label{eq:hamiltonian}
\mathcal{H} \approx \frac{m_a^2 f_a^2}{2}\left( \frac{\theta'^2}{\hat\eta_c^{n} \hat\eta^2} + \theta^2 \right) \,,
\end{equation} 
when all modes in the simulation are non-relativistic and the field values are small.

\subsection{Oscillons}
Large overdensities right after the QCD phase transition are caused by oscillons. Oscillons are, in contrast to strings and domain walls, not topological defects but arise due to non-linearities in the equation of motions, forming at locations where the the axion self-interaction dominates the Hubble friction. As a result, the first oscillons form at the location of collapsed strings and domain walls, where the axion remains excited and reaches large field values. However, at later times, oscillons are observed forming throughout the simulation box. The dynamics of the oscillons are highly non-trivial, especially as the axion self-interaction increases in strength with the growing axion mass. 

Oscillons decrease in size over time following the oscillation wavelength $\sim$$m_a(T)^{-1}$, as axions in the core are relativistic. Good spatial resolution is therefore needed to resolve them. In order to
study their behavior we perform a 2D (two spatial dimensions, one time) simulation using the same simulation setup in the PQ- and QCD-epoch as in 3D. We find that there is no qualitative difference between 2D and 3D simulations regarding oscillons, but going to 2D allows us to increase the spatial resolution to $4096^2$ grid sites and to subsequently increase $\hat\eta_c$.

\begin{figure*}[htb]
\includegraphics[width=1\textwidth]{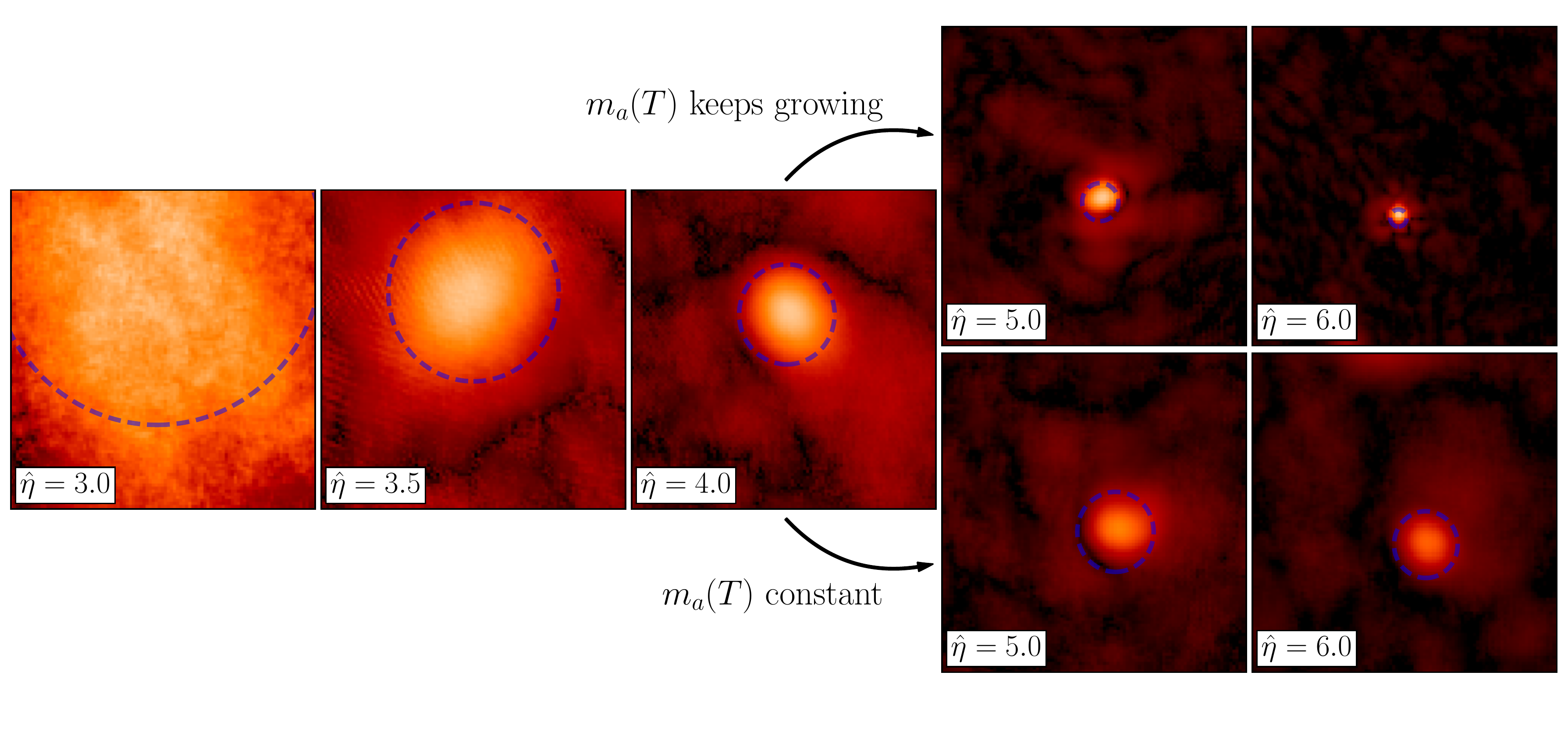}
  \caption{Illustration of an oscillon ($\log(\rho/\bar\rho)$) at different times in a 2D simulation. Two scenarios are considered with different truncation points of the mass growth, $\hat\eta_c=4.0$ and $\hat\eta_c=6.0$. The three left panels are identical in both scenarios, while the two top right panels are for $\hat\eta_c=6.0$, and the two bottom panels are for $\hat\eta_c=4.0$. The radius of the oscillon is proportional to the oscillation frequency $~\sim m_a(T)^{-1}$ (circles of that radius are shown in dashed blue) and as such is decreasing over time.  The oscillon central density slowly dissipates after the mass growth ends, as seen in the bottom right panels for $\hat \eta_c = 4.0$.
  }
  \vspace{0.3cm}
  \label{fig:oscillon}
\end{figure*}

We illustrate the evolution of an oscillon in Fig.~\ref{fig:oscillon}. Two scenarios are considered with different truncation points of the mass growth, $\hat\eta_c=4.0$ and $\hat\eta_c=6.0$. Note how the radius of the oscillon decreases as long as $m_a(T)$ is increasing.  The circles in~\ref{fig:oscillon} have radius $m_a(T)^{-1}$, and the oscillon cores are seen to track this scale.
Subsequently, if the mass growth is truncated at $\hat\eta_c=4.0$, the radius of the oscillon is constant as well.  When the mass growth is cut-off, the density contrast at the core of the oscillon slowly decreases over time and the oscillons dissipate, as can be seen
in the two lower right panels in Fig.~\ref{fig:oscillon}.

\subsection{Calculating the Axion Relic Abundance}

To calculate the axion DM abundance as a function of $m_a$, we first need to understand the relationship between the mass cutoff conformal time $\hat \eta_c$ and the decay constant $f_a$. 
Here we use the relation $T_1 / \hat\eta_c = T_c$, with $T_c \approx 100$ MeV. This allows us to solve for $f_a$ in terms of $\hat\eta_c$. The energy densities are calculated from the axion field and its derivatives  according to~\eqref{eq:hamiltonian} after numerically evolving until $\hat\eta = 7$, then analytically evolving until $\hat\eta_{\rm MR} = 10^6$, at which point the contribution of the gradient term to the energy density is negligible.  As a side note, our definition of $\hat\eta_{\rm MR}$ actually puts us at slightly earlier times than global matter-radiation equality.  This us because matter-radiation equality is, locally, reached earlier for the largest overdensities and because we want to make sure that gravitational interactions can be neglected.
  In particular, note that the temperature corresponding to $\hat \eta_{\rm MR}$ is given by $T_{\rm MR} = T_c \eta_c / \hat \eta_{\rm MR}$.  For our most realistic simulation with $\eta_c = 3.6$ this corresponds to $T_{\rm MR} \approx 0.5$ keV.  However, if we reinterpret the final state for a more realistic axion with $m_a \approx 25$ $\mu$eV, which has a higher $\eta_c$, then $T_{\rm MR} \approx 4$ keV.  In practice, though, the exact value of $T_{\rm MR}$ is not important because by these temperatures the proper motions in the axion field are frozen out and the field is thus not evolving non trivially.
  As a consequence our results (both for the DM density and for the spectrum of overdensities) are not sensitive to small (or even relatively large) changes to the exact value of $\hat \eta$ that we evolve to.  

\begin{figure*}[htb]
\includegraphics[width=.4\textwidth]{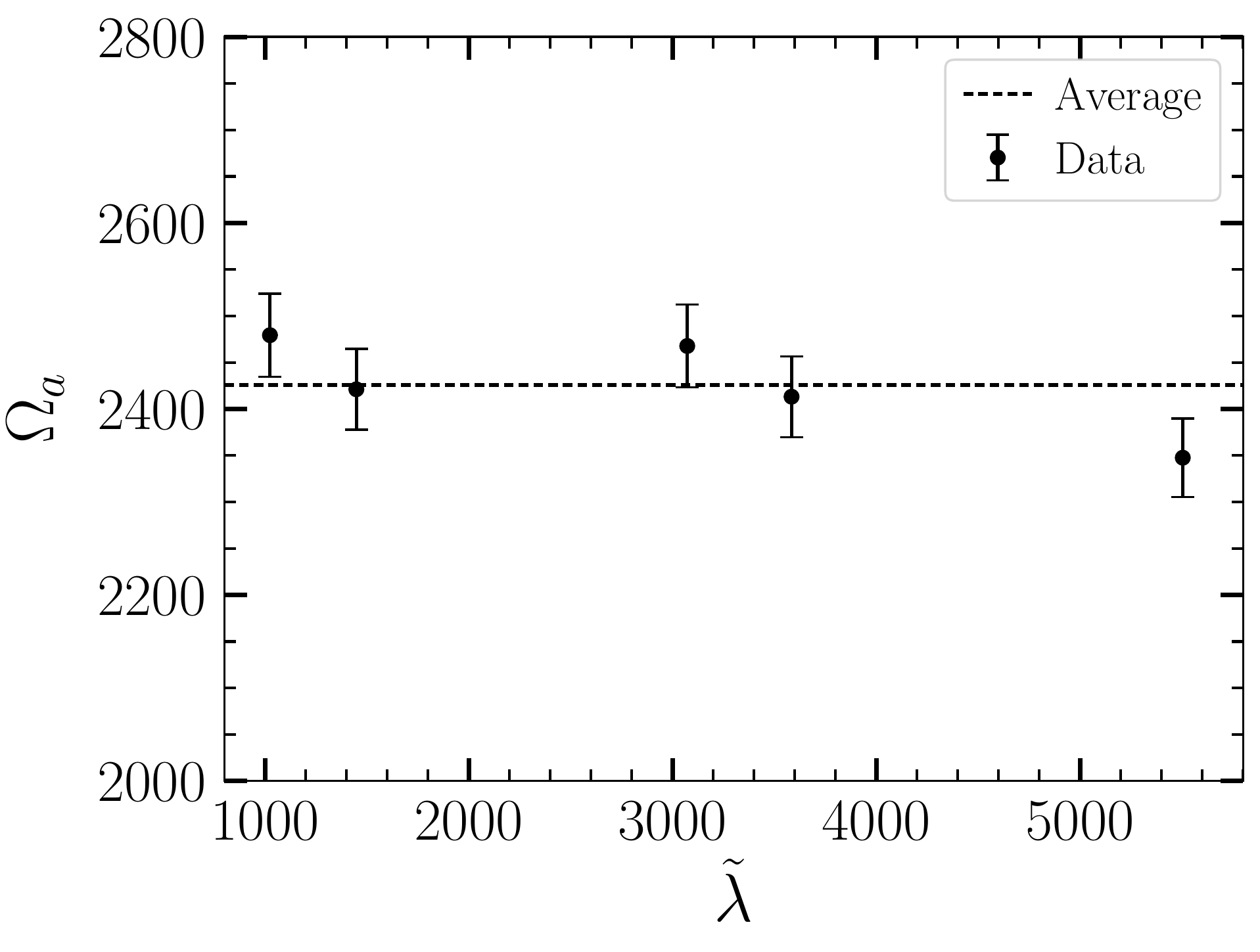}
  \caption{Our results for the DM density today $\Omega_a$, inferred at $\hat \eta_{\rm MR}$, from simulations at different values of $\tilde \lambda$ for our most realistic $\hat \eta_c$: $\hat \eta_c = 3.6$.  The uncertainties are the inferred statistical uncertainties arising from the spread in the DM density determinations as a function of $\tilde \lambda$.  No trend is discernible for the dependence of $\Omega_a$ on $\tilde \lambda$, above the statistical noise.}
  \vspace{0.3cm}
  \label{fig: DM-lambda}
\end{figure*}

Note that we present our results in terms of the DM density fraction today $\Omega_a$, which is defined as the ratio of the average energy density today in DM relative to the observed critical energy density.
We compute statistical error bars at each value of $f_a$ from the variance as a function of $\tilde \lambda$ at fixed $\hat\eta_c$.  We note that no trend is visible in the data for the dependence of $\Omega_a$ on $\tilde \lambda$, as is shown in Fig.~\ref{fig: DM-lambda}.  
The statistical noise is inferred from the spread in $\Omega_a$ values, which are determined from the output at $\eta_{\rm MR}$, between different $\tilde \lambda$.  The observed variations are consistent with the expected noise from Poisson counting statistics due to having a finite number of overdensities within the simulation box.  

\begin{figure*}[htb]
\includegraphics[width=.75\textwidth]{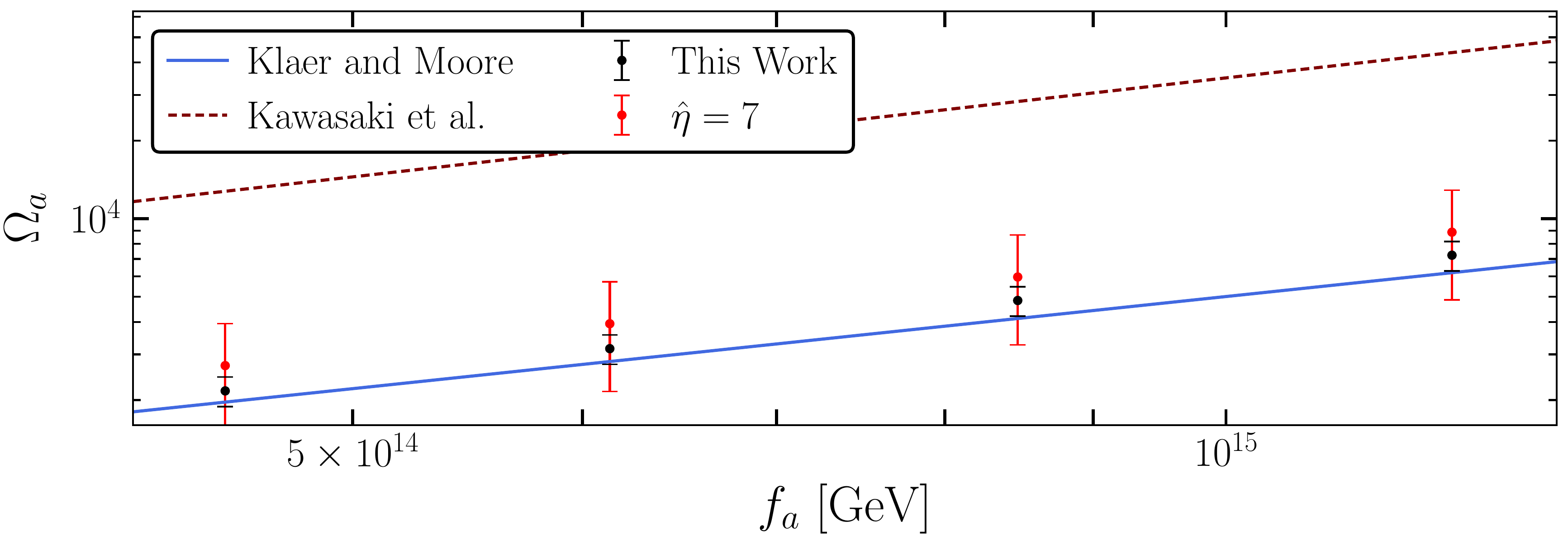}
  \caption{A comparison of the predictions for the relic abundance of axions dark matter as a function of $f_a$ obtained in \cite{Kawasaki:2014sqa} (Kawasaki et al.) and  \cite{Klaer:2017ond} (Klaer and Moore) with the simulation results realized in this work. Error bars are combined statistical and correlated systematic errors, with the former dominating at $\hat\eta=7$ due to large field gradients and the latter at $\hat\eta=\hat\eta_{MR}$.}
  \vspace{0.3cm}
  \label{fig:Relic}
\end{figure*}

In Fig.~\ref{fig:Relic} we show our results for $\Omega_a$ as a function of $f_a$, compared to earlier predictions in  \cite{Kawasaki:2014sqa} and  \cite{Klaer:2017ond}.
For reference, we also include predictions for the relic abundance based on the field value and the time derivative at $\hat\eta = 7$.  Here it is less straightforward to determine the DM axion abundance, relative to taking the results at $\hat \eta_{\rm MR}$, as some of the modes in the simulation are still relativistic.  This introduces an additional systematic uncertainty, since the field is not completely red-shifting like radiation at this time.  For these reasons it is important to evolve the field until it is completely non-relativistic before measuring the DM density.

Because the ratio of the axion mass density to entropy density is constant after the axions have become non-relativistic and the number of axions is conserved, we can redshift our energy density from our matter-radiation equality $\hat \eta_{\rm MR}$ to today.  Then, we compare this energy density to the most up-to-date measurement of the average DM density in the Universe today $\rho_{\rm DM} = 33.5 \pm 0.6$ $M_\odot / {\rm kpc}^{3}$~\cite{Aghanim:2018eyx}.  
Note that we have propagated all cosmological uncertainties other than those on $N_{\rm eff}$, which we have fixed to the Standard Model value. These cosmological uncertainties introduce an approximately $3\%$ correlated uncertainty across the results of our simulations. We additionally have an approximately $8\%$ uncertainty due to our assumption of fixed $g_*$, which is examined in greater detail later in this Supplement. These uncertainties are the dominant ones in our results, and we emphasize that they have not been typically considered in determinations of the DM axion mass.
From the $\Omega_a$ data, for the various $f_a$ values simulated, we may extrapolate to predict the $f_a$ for an axion which produces the observed DM relic abundance by fitting a simple power law relation of the form
\begin{equation}
\Omega_a(f_a) = c_1 \cdot f_a^\alpha \,,
\label{eq:Omega}
\end{equation}
as discussed in the main body of this work.  Note that we expect $\alpha =  (6 + n)/(4+n) $, where $n$ is the index of the axion mass growth.  We assume this scaling is valid to make our estimate for the $m_a$ that gives the correct DM abundance.  The relation between $\alpha$ and $n$ is expected to arise for the following reason.  Let us estimate the axion DM density from an axion with a constant initial misalignment angle $\theta_i$. The present-day axion abundance as produced by the misalignment mechanism can be estimated by 
\begin{equation}
\rho_a(T_0) = \rho_a(T_3) \frac{m_a(T_0)}{m_a(T_3)}\frac{g_*(T_0)T_0^3}{g_*(T_1) T_3^3} \,,
\end{equation}
where $T_0$ is the present-day temperature, $T_3$ is the temperature at which the axion began to oscillate ($m_a(T_3) = 3 H(T_3)$), and $g_*(T)$ the number of effective degrees of freedom at temperature $T$.

The initial axion abundance $\rho_a(T_1)$ is given
\begin{equation}
\rho_a(T_1) = \frac{m_a(T_1)^2 f_a^2 }{2}\theta_i^2 \,,
\end{equation}
 Anharmonicity factors can be included, but have no temperature or $f_a$ dependence. The temperature $T_3$ depends on $f_a$ through the relation $T_3 \propto f_a^{-2/(4+n)}$. Substituting these relations in and keeping only terms which depend on $f_a$, we have 
\begin{equation}
\rho_a(T_0)  \propto f_a^{(6+n) / (4+n)} \frac{g_*(T_0)}{g_*(T_3)} \,.
\end{equation}
We thus expect the relic abundance to scale with $f_a$ like $\rho_a \propto f_a^{(6 + n) / (4 + n)}$. 
Note that the DM abundance from string and domain wall production is calculated similarly in \cite{Kawasaki:2014sqa}, and although our results are not consistent with those presented in that work, the abundance calculation they present proceeds similarly, yielding string and domain wall production that scale like $f_a^{(6+n) / (4+n)}$ as well. 

On the other hand, we may also calculate the the $m_a$ that gives the correct DM abundance by using our fit value for $\alpha$, as defined in~\eqref{eq:Omega}, instead of the theoretical value.  Doing so leads to a slightly lower $m_a$ estimate, as described in the main text.

\subsection{Tests of the Overdensity Field Gaussianity}
\begin{figure*}[htb]
\includegraphics[width=.75\textwidth]{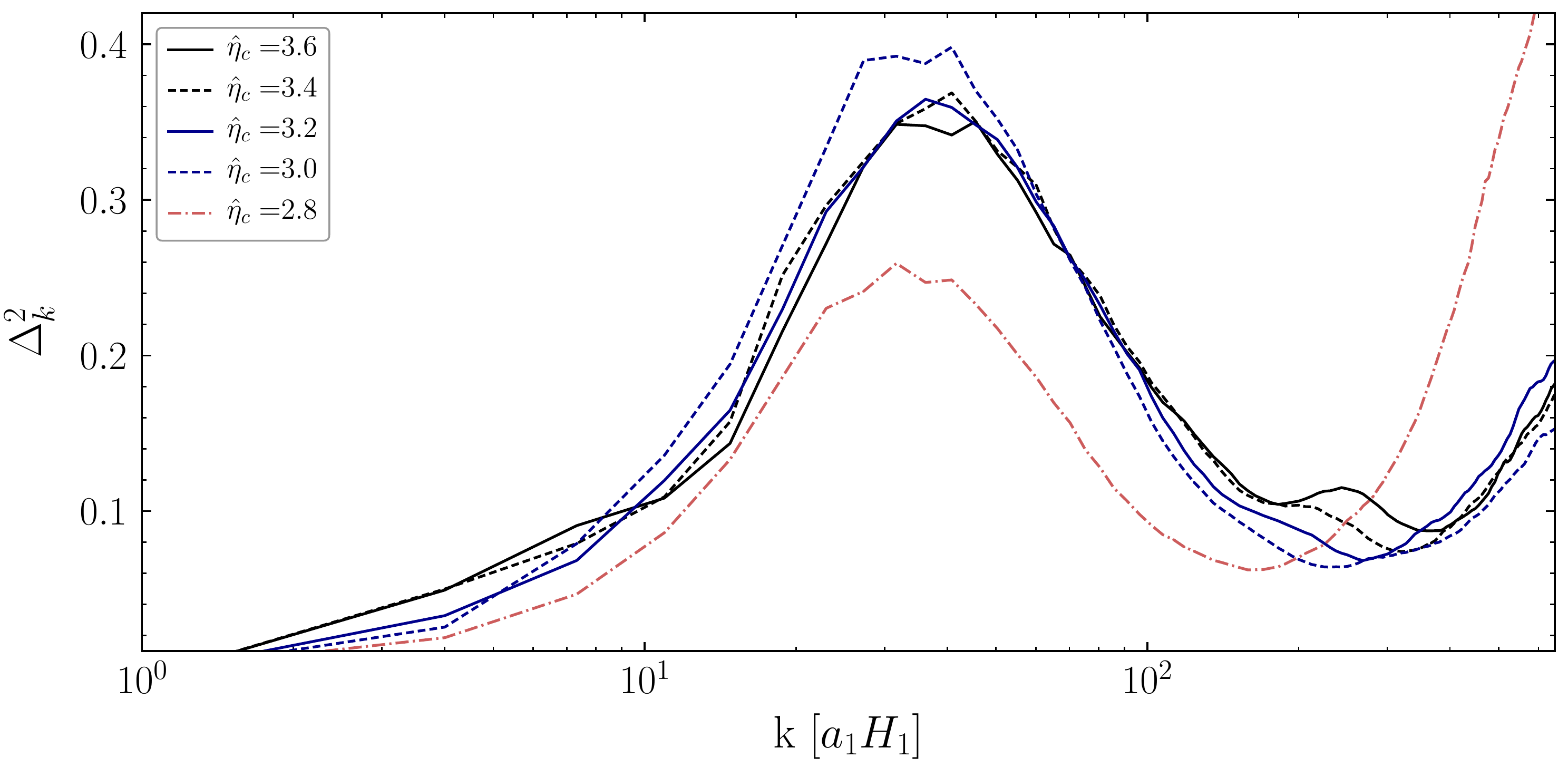}
  \caption{A comparison of the power spectra realized in simulations for $\tilde \lambda = 5504$ for different choices of $\hat\eta_c$. New features in the power spectrum emerge as we push to larger values of $\hat\eta_c$, and we cannot exclude the possibility that further features would emerge were we to simulate with a greater value for the cutoff.  On the other hand, the power-spectrum is highly non-Gaussian at small scales, so the distribution $\Delta_k^2$ alone is not adequate for understanding the small-scale nature of the overdensity field.}
  \vspace{0.3cm}
  \label{fig:PowerSpectra}
\end{figure*}

In typical cosmological contexts, overdensity fields are treated under the assumption that they are Gaussian random fields. For a real-space Gaussian field, we may Fourier transform the field and find that the squared magnitude of each mode is independently exponentially distributed with mean set by the power-spectrum and with the phase of each mode independently uniformly distributed on $[0, 2\pi)$ \cite{doi:10.1137/1.9780898718980}. For reference, in Fig.~\ref{fig:PowerSpectra} we show our power spectra $\Delta_k^2$ at fixed $\tilde \lambda$ across our various choices for $\hat\eta_c$.  Note that we construct the power spectra from the fields that have been evolved until $\hat\eta = \hat \eta_{\rm MR}$. However, as we demonstrate below, the power spectrum fails to accurately describe the overdensity field we realize in our simulations because the field is highly non-Gaussian at small scales. As a result, standard tools for predicting structure formation that rely upon an underlying Gaussian overdensity field, such as the Press-Schechter formalism, cannot be applied to predict the spectrum of structures that form from the overdensities in the axion field, at least on the very smallest scales. 
 
First, we note that the largest field values taken within the overdensity fields at the state realized by the analytic evolution until $\hat\eta = \hat \eta_{\rm MR}$ are $\mathcal{O}(10)$, whereas the minimum value the overdensity field can take is $-1$ by construction. This is trivially incompatible with the interpretation of the overdensity field as a Gaussian random field, which would have symmetric variance about its mean of $0$. For our overdensity fields to realize $\mathcal{O}(10)$ maxima with $-1$ as a construction-imposed minimum, there must exist considerable phase-correlations between Fourier modes, contrary to the uncorrelated phases of a Gaussian random field.

\begin{figure*}[htb]
\includegraphics[width=1.\textwidth]{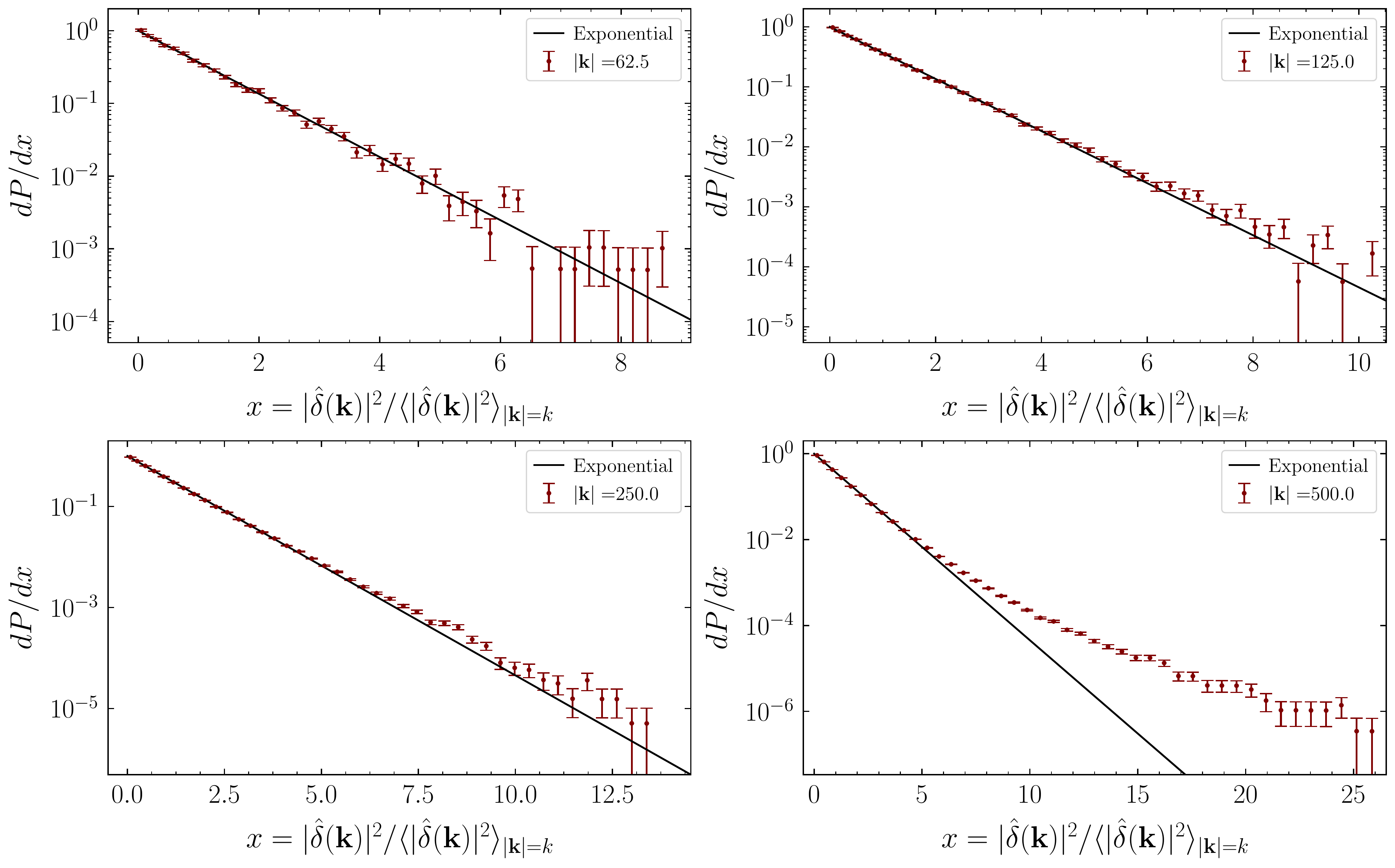}
  \caption{A comparison of the distribution of the squared magnitudes of Fourier components for four different fixed reference momentum $k$.  The expected exponential distribution for a Gaussian field is also indicated.  While the distributions are Gaussian at large scales, they become increasingly non-Gaussian at small scales.  These distributions were constructed from our most realistic  simulation with $\tilde \lambda = 5504$ and $\eta_c = 3.6$.}
  \vspace{0.3cm}
  \label{fig:PowerDist}
\end{figure*}

We also may inspect the distribution of power at each mode in the Fourier transformed overdensity field. If the overdensity field were Gaussian, then the power in each mode would be exponentially distributed with mean set by the value of the mean power spectrum. To test this, we plot the probability distribution $dP/dx$ of $x = |\hat \delta(\mathbf{k})|^2 / \langle|\hat \delta(\mathbf{k})|^2 \rangle_{|\mathbf{k}| = k}$, with $\hat \delta({\bf k})$ the Fourier transformed overdensity field at momentum ${\bf k}$, as measured in the final states of our field at $\hat\eta= \hat \eta_{\rm MR}$. 
We compare the observed distributions with the expected Gaussian random field assumption of an exponential distribution with unit mean in Fig.~\ref{fig:PowerDist}.  Dramatic deviations from the expected behavior are observed for large $|\mathbf{k}|$.  We stress, however, that in addition to these distributions departing from the expected exponential distributions, the real and imaginary components across modes are also highly phase correlated on small scales. 

\subsection{Minihalo Mass Spectrum}

In this subsection we give additional details and results for the minihalo mass and density spectrum.
In addition to the technical difficulties associated with a non-Gaussian overdensity field, computational limitations prevent us from performing realistic simulations of $f_a \sim 10^{11} \, \mathrm{GeV}$ axions, which would require us to simulate until $\hat\eta_c \approx 15$. We instead interpret our simulation results at smaller $\hat\eta_c$ in appropriate units to rescale these results to the target $f_a \approx 2\times10^{11} \, \mathrm{GeV}$.  We do so with the following methods. The total axion mass contained within some set of grid-sites in our simulation can be computed from the Hamiltonian as
\begin{equation}
M_{\rm tot} = a(\hat\eta)^3 \int d^3 x \mathcal{H} \approx a(\hat\eta) \sum (\Delta x)^3 \mathcal{H} = \left(\frac{a(\hat\eta)^3 \Delta \bar x}{a_1 H_1}\right)^3\sum \mathcal{H} =  \left(\frac{\hat\eta \Delta \bar x}{H_1}\right)^3 \sum (1 + \delta) \bar \rho \,,
\label{eq:MHmass}
\end{equation}
where $\bar \rho$ is computed by the average of our Hamiltonian in~\eqref{eq:hamiltonian} in the simulation box. We calculate $H_1$ from $T_1$ based on our choice of $f_a$, then rescale $\bar \rho$ to the value of the axion energy density at the time $\hat\eta$ such that the correct relic abundance is realized today. In this manner, we aim to rescale all dimension-full quantities related to $f_a$ to our target $f_a$.  In particular, we rescale the DM density $\bar \rho$ to give the correct DM density realized in our Universe, and we also rescale the minihalo masses by the factor $\propto (a_1 H_1)^{-3}$ appearing in~\eqref{eq:MHmass} to those for the target $f_a$.

\begin{figure*}[htb]
\includegraphics[width=.66\textwidth]{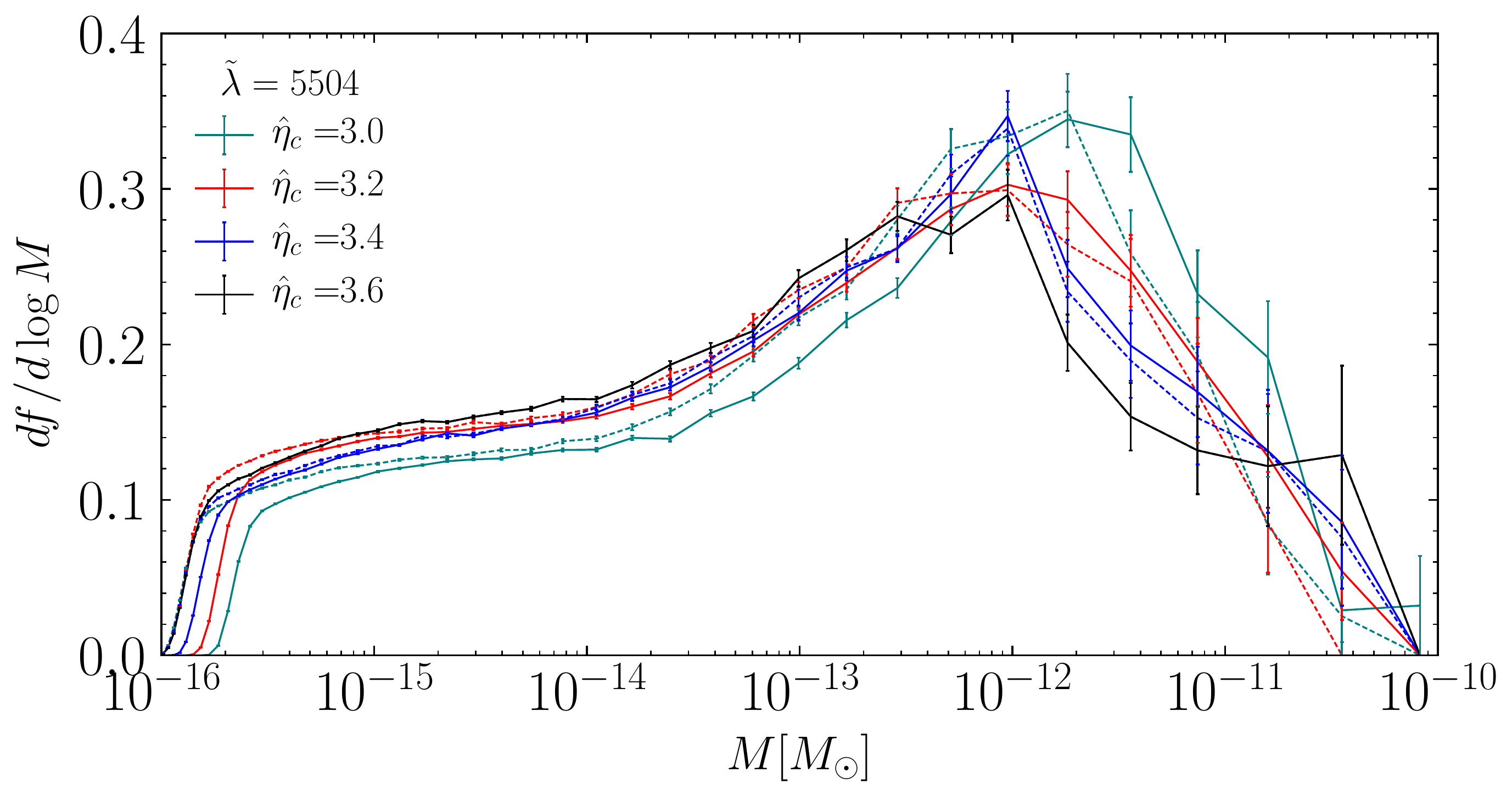}
  \caption{Comparison between differential mass fractions as a function of the minihalo mass $M$ from our simulations at different $\hat \eta_c$.  In this plot we have rescaled the minihalo masses such that we achieve the correct DM density $\bar \rho$ observed in the Universe, but for the solid curves we have not applied the Hubble volume rescaling factor to reach our target $f_a$.  However, the dashed curves do have the Hubble volume rescaling factor included, but here we take our target $f_a$ to be that corresponding to our most realistic simulation with $\hat \eta_c = 3.6$.  The difference between the dashed mass functions and the solid black mass functions gives a sense of the systematic uncertainty introduced by applying the naive mass rescaling factors instead of simulating with the correct value of $\hat \eta_c$ ($f_a$).   }
  \vspace{0.3cm}
  \label{fig:M-rescale}
\end{figure*}
We illustrate the rescaling procedure in Fig.~\ref{fig:M-rescale}.  In that figure we show the differential mass distribution of minihalos $df / d \log M$ as a function of minihalo mass $M$.  These mass distributions have been rescaled such that $\bar \rho$ matches the actual DM density.  However, the solid curves do not have the $a_1 H_1$ Hubble volume rescaling included.  The dashed curves, on the other hand, apply the Hubble volume rescaling factor but for a target $\hat \eta_c$ of $\hat \eta_c = 3.6$, which is that corresponding to the black curve.  Clearly there are still differences between the rescaled dashed curves and the black curve, which tells us that there are dynamical effects that arise from changing $\hat \eta_c$ that are not captured by the simple rescaling.  This should not be too surprising considering that {\it e.g.} the mass growth affects the oscillon stability, which determines the high-mass part of the distribution.  In our work we rescale the mass function to the target $f_a$ as described above, but it is important to keep in mind that this almost certainly results in a systematic uncertainty from the fact that we do not capture the full oscillon dynamical range in doing so.  Also note that all of the mass functions abruptly drop off at low halo masses.  This is due to our resolution limit on the finite lattice.  We also cannot rule out the possibility that the low-mass tail continues down to much smaller masses.

\begin{figure*}[htb]
\includegraphics[width=0.49\textwidth]{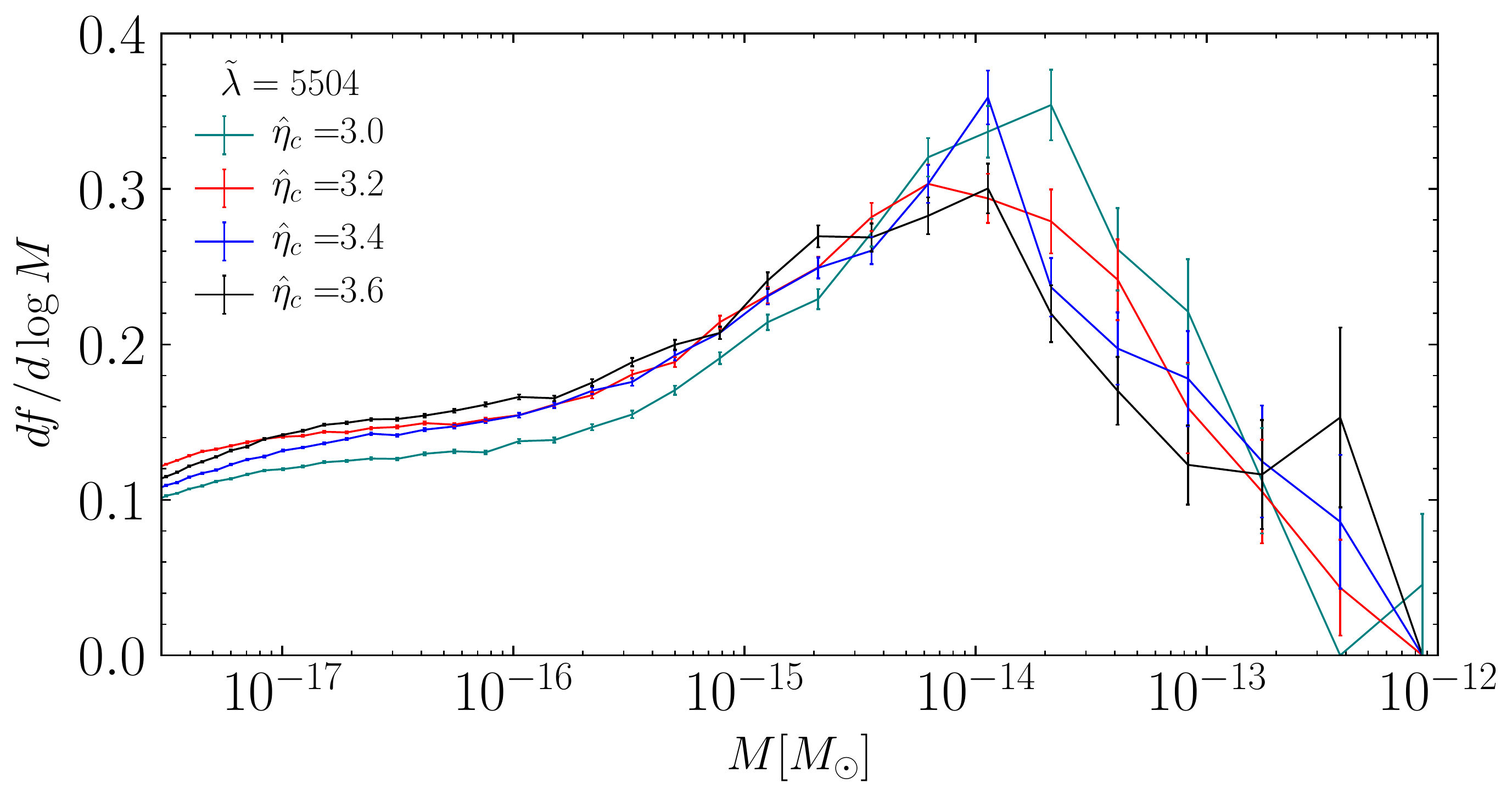}
\includegraphics[width=0.49\textwidth]{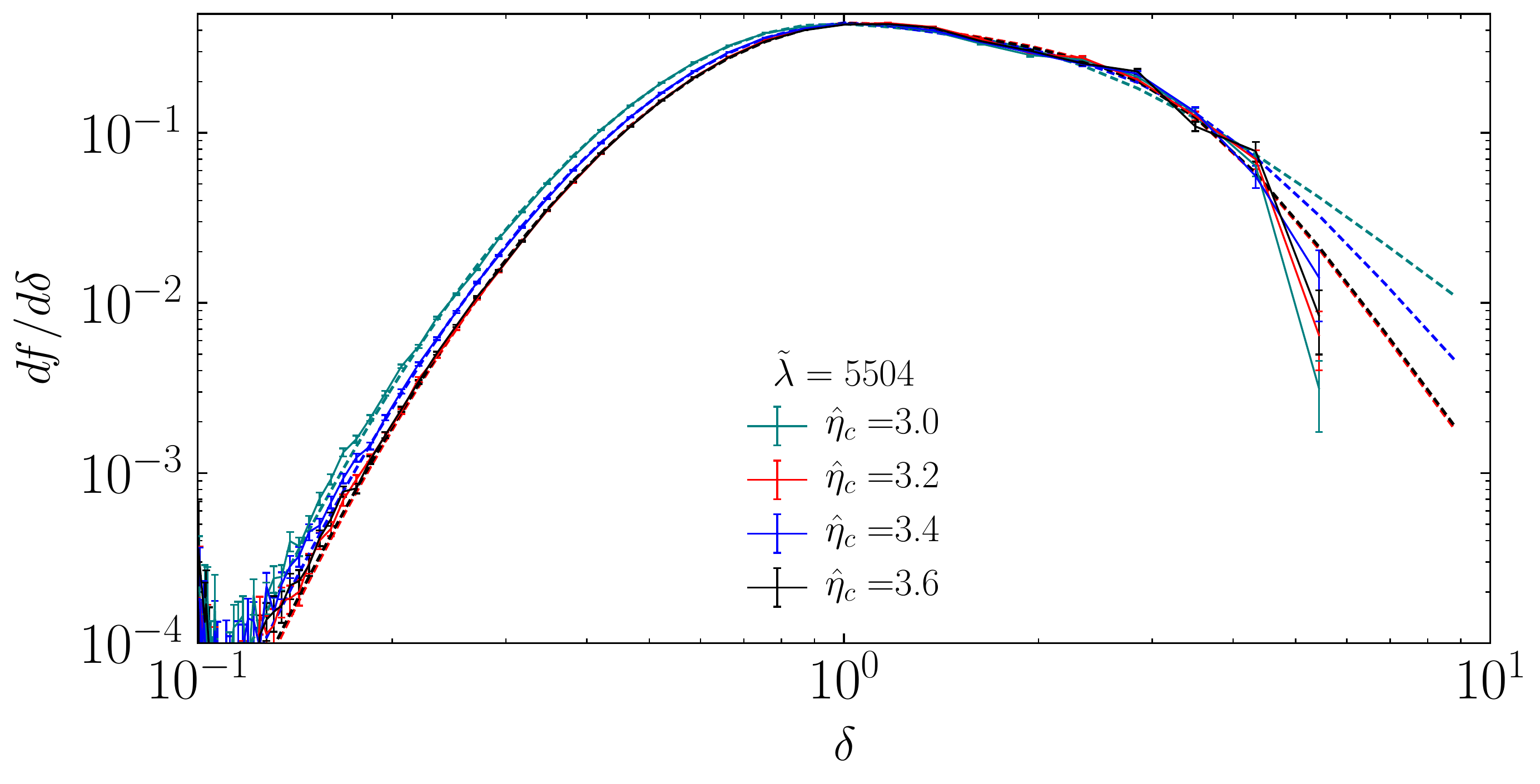}\\
\includegraphics[width=0.49\textwidth]{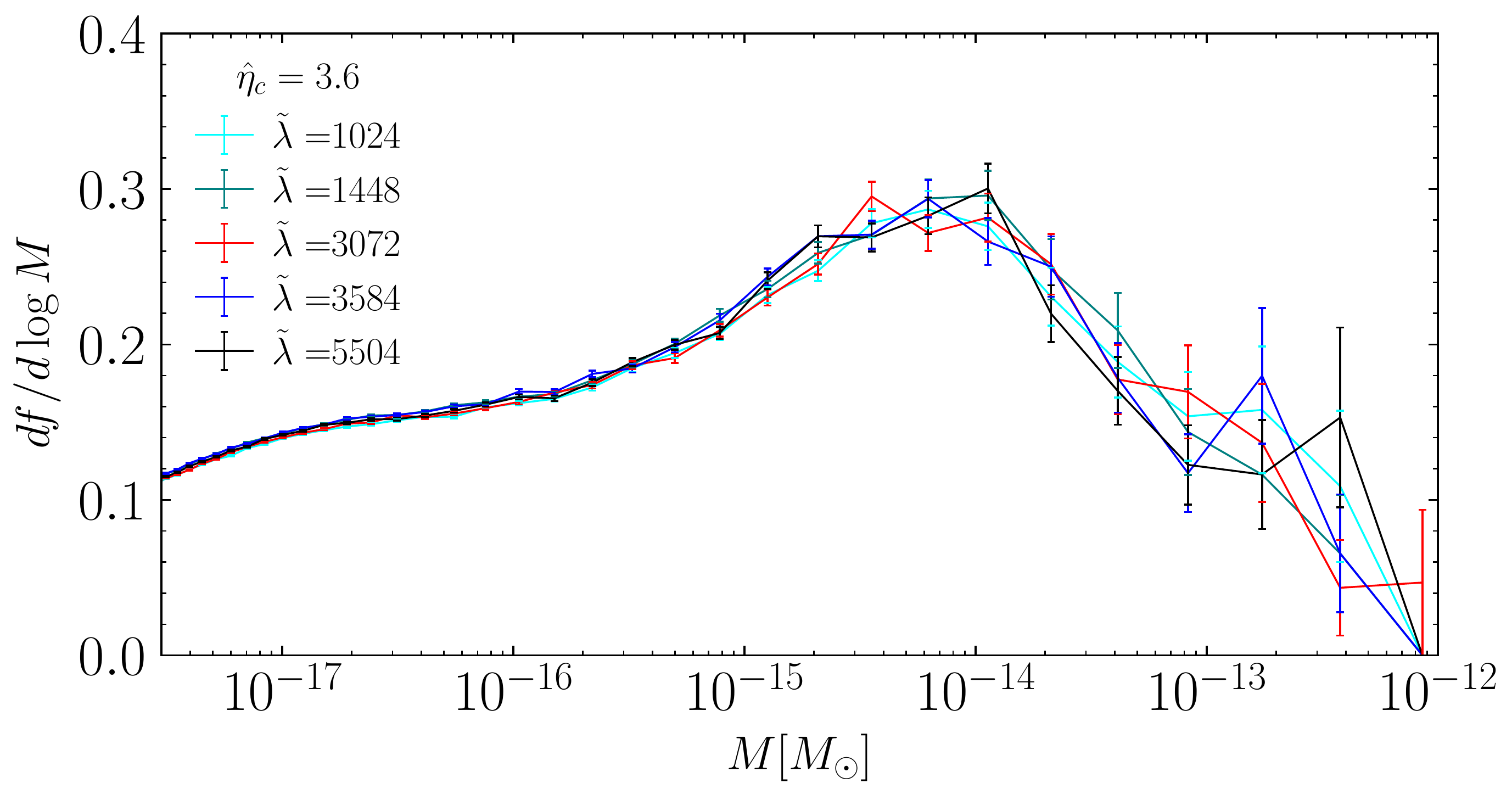}
\includegraphics[width=0.49\textwidth]{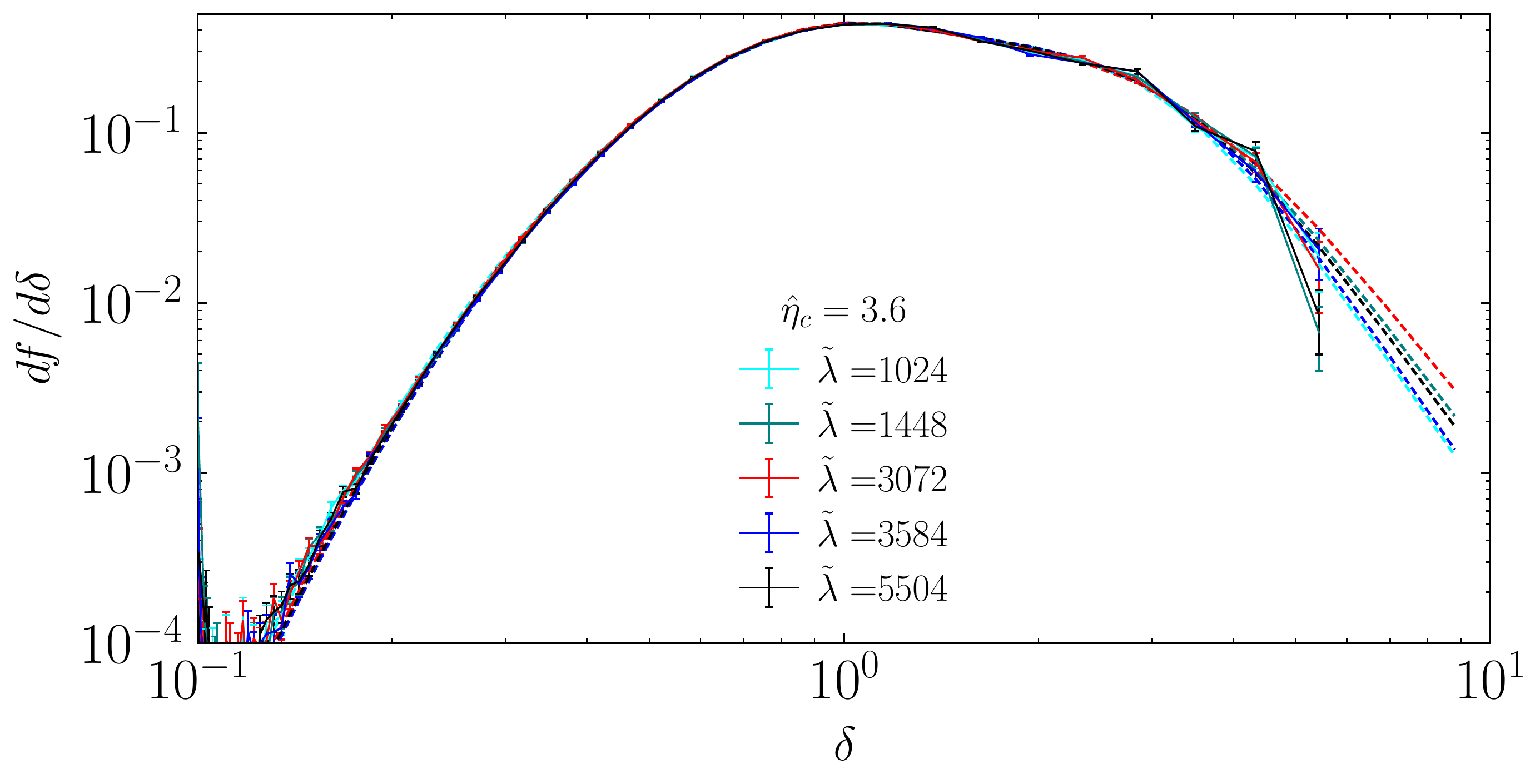}
  \caption{Comparison between differential mass fractions as a function of the concentration parameter $\delta$ and minihalo mass $M$ for different $\hat\eta_c$ and $\tilde\lambda$ at $\hat\eta=\hat\eta_\text{MR}$. Error bars are statistical. Shown as dotted lines is a fit to the $df/d\delta$ curves as described in the text. We do not extend the $df/d\log M$ curves to lower masses as we are unable to resolve those properly.}
  \vspace{0.3cm}
  \label{fig:MDProjections}
\end{figure*}

We compare the single-differential mass fractions for different values of $\hat\eta_c$ and $\tilde\lambda$ as a function of $\delta$ and $M$ 
in Fig.~{\ref{fig:MDProjections}}.  Note that here we have applied the rescaling factors for the masses to our true target $f_a$, which is that which gives the correct DM density.  First of all we note that there is no dependence on $\tilde\lambda$ visible in our parameter range within than statistical scatter. As for the differential distribution as a function of $\delta$, there is also no clear dependence on $\hat\eta_c$ visible. 
 The only place where a clear dependence on $\hat\eta_c$ is visible is in the mass fraction as a function of $M$. Here, the peak values shift to smaller masses upon increasing $\hat\eta_c$, even after including the rescaling factors.  

It is useful to have an approximate analytic formula for the differential mass fraction.  We find that the differential mass fraction as function of $\delta$ can be accurately described by a Crystal Ball function based on a generalized Gaussian
and a power-law high-end tail together with a suppression factor at high-$\delta$:
\begin{equation}
\frac{df}{d\delta} = \frac{A}{1+\left(\frac{\delta}{\delta_F}\right)^S}
\begin{cases}
      e^{-\left[\ln{\left(\frac{\delta}{\delta_G}\right)/\sqrt{2}\sigma}\right]^d} & \text{for } \ln{\left(\frac{\delta}{\delta_G}\right)}\leq\sigma\alpha\\
      B\left[C+\frac{1}{\sigma}\ln{\left(\frac{\delta}{\delta_G}\right)}\right]^{-n} & \text{for }\ln{\left(\frac{\delta}{\delta_G}\right)}>\sigma\alpha
    \end{cases} \,.
\end{equation}
The parameters $B$ and $C$ are given by 
\begin{equation}
B=e^{-\left(\frac{|\alpha|}{\sqrt{2}}\right)}\left[\left(\frac{\sqrt{2}}{|\alpha|}\right)^d\frac{|\alpha|n}{d}\right]^n,\qquad 
C=|\alpha|\left[\left(\frac{\sqrt{2}}{|\alpha|}\right)^d\frac{n}{d}-1\right] \,,
\end{equation}
and they are chosen such that $df/d\delta$ and its first derivative are continuous. $A$ is not a free parameter as $\int_{0}^\infty d\delta (df/d\delta) =1$ must hold. The fit parameters from our most realistic simulation with $\hat\eta_c=3.6$ and $\tilde\lambda=5504$ are given by
\begin{center}
\begin{tabular}{llll}
$\sigma=0.448\pm0.008\qquad\qquad$	& $n =115\pm8$		& $\delta_G=1.06\pm0.02\qquad\qquad$	& $S=4.7\pm1.6$ \\
$         d=1.93\pm0.02$	& $\alpha =-0.21\pm0.07\qquad\qquad$	& $\delta_F=3.4\pm1.2$ \,. 		&      
\end{tabular}
\end{center}

\begin{figure*}[htb]
\includegraphics[width=0.66\textwidth]{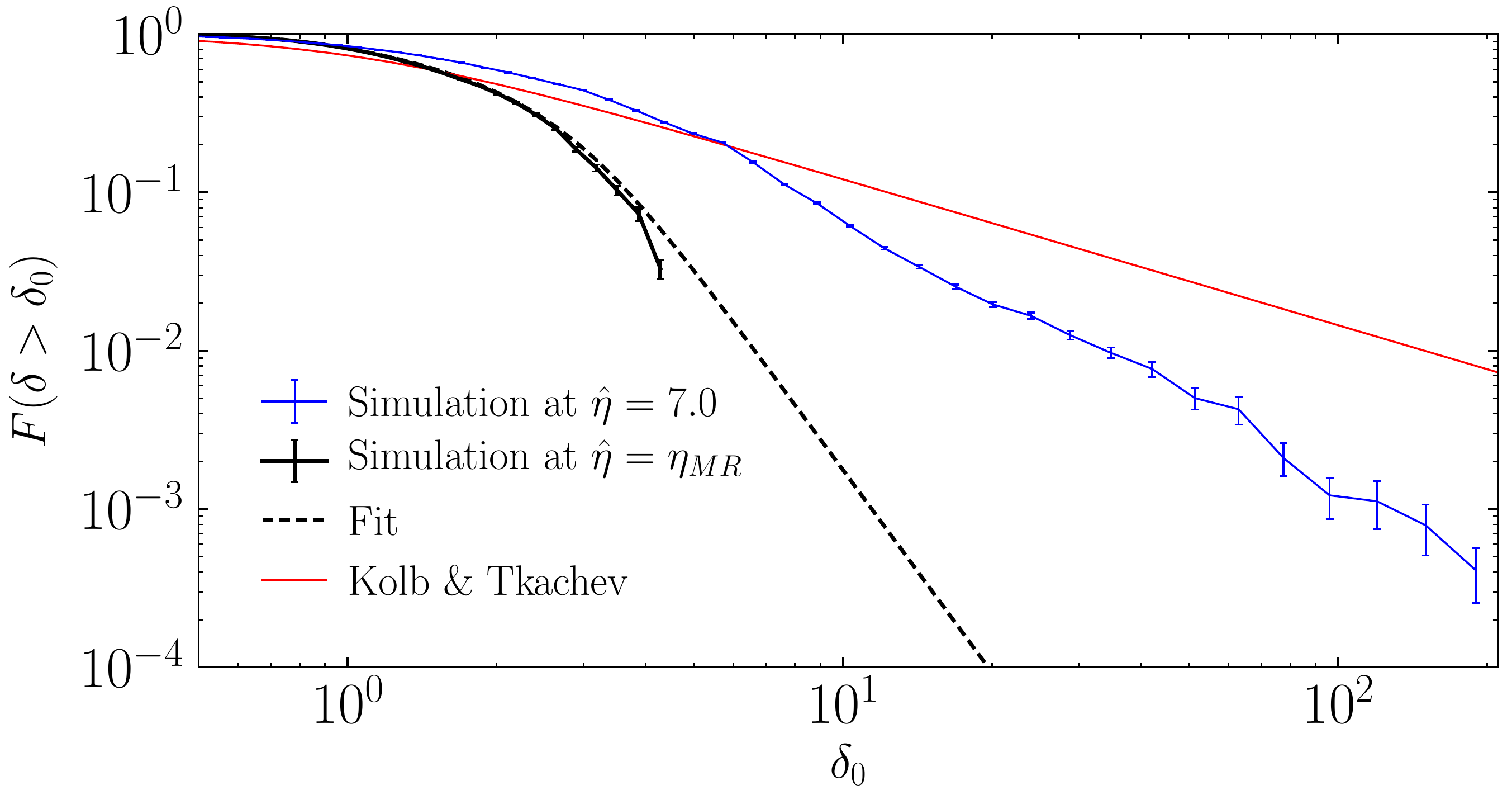}
  \caption{Comparison between cumulative mass fractions, defined in the text, for our simulation at $\hat\eta=7$ (solid blue) and $\hat\eta_\text{MR}$ (solid black). We use our fit to the differential mass fraction $df/d\delta$ to extrapolate to high $\delta_0$ for our $\hat\eta_\text{MR}$ data (dotted black). Error bars are statistical. We compare our results to those from Kolb and Tkachev~\cite{Kolb:1995bu} obtained at $\hat\eta=4$ by using the fit to their data presented in~\cite{Fairbairn:2017sil} (red curve).}
  \vspace{0.3cm}
  \label{fig:Kolb}
\end{figure*}

This fit allows us to make a precise comparison with previous work by Kolb and Tkachev~\cite{Kolb:1995bu}. We present in Fig.~\ref{fig:Kolb} the 
cumulative mass fraction that is in overdensities larger than $\delta_0$,
\begin{equation}
F(\delta>\delta_0) = \int_{\delta_0}^{\infty}\frac{df}{d\delta} d\delta.
\end{equation}
Unsurprisingly, we find considerably less mass in highly concentrated overdensities relative to~\cite{Kolb:1995bu}. Whereas~\cite{Kolb:1995bu} predicts roughly $10\%$ of the mass is in overdensities with $\delta=10$ or more, we find a similar result only when using the simulation output at $\hat\eta=7$. Once evolved to matter-radiation equality, that percentage falls to $\sim$$0.1$\%. 

\section{Equations of Motion for Varying Relativistic Degrees of Freedom}

In this section we investigate the systematic effect on our results from the assumption of fixed $g_*$.
In truth, the value of $g_*$ is not fixed at $g_* \approx 81$ but instead varies rather sharply in the temperature range of interest; it varies from as large as roughly $100$ to as little as roughly $10$ around the time of the QCD phase transition. This does not represent a dire shortcoming of our simulation procedure, however, as varying $g_*$ should only nontrivially affect the dynamics of the axion field during times when axion number density is not a conserved quantity.  By $\hat\eta \approx 3$ most of the field has become linear, except for the isolated oscillon configurations, which means that the axion number density is mostly conserved at this time and beyond.  The variation in $g_*$ before $\hat \eta \approx 3$ for our target $f_a$ is relatively minor.
 To quantify this impact, however, we perform 2D simulations accommodating the varying $g_*$.

With a change of variable we may rewrite the axion equations of motion, in the two-field formalism during the QCD epoch, as 
\begin{gather}
 \phi_1'' + \left(\frac{R_1 \ddot R}{\dot R^2} + \frac{3 }{\eta} \right)\phi_1' - \frac{R_1^2 \dot R_1^2}{R^2 \dot R^2}\nabla^2 \phi_1 + \frac{\dot R_1^2}{\dot R^2}\left[  \tilde \lambda \phi _1  \left(\phi _1^2+\phi _2^2 - 1\right) -\frac{m_a(T)^2\phi _2^2 }{ H_1^2 \left(\phi _1^2+\phi _2^2\right){}^{3/2}} \right] = 0 \\ 
\phi_2'' + \left(\frac{R_1 \ddot R}{\dot R^2} + \frac{3}{\eta} \right)\phi_2' - \frac{R_1^2 \dot R_1^2}{R^2 \dot R^2} \nabla^2 \phi_2 +\frac{\dot R_1^2}{\dot R^2}\left[\tilde \lambda  \phi _2 \left(\phi _1^2+\phi _2^2 - 1\right) + \frac{m_a(T)^2 \phi _1 \phi _2}{H_1^2\left(\phi _1^2+\phi _2^2\right)^{3/2}}\right]= 0 \,,
\end{gather}
where we define $\tilde \lambda = \lambda f_a^2 /H_1^2$ as before. Citing standard references \cite{Kolb:1990vq}, we have
\begin{gather}
H \approx 1.660 g_*(T)^{1/2} \frac{T^2}{m_{Pl}} \\
t \approx 0.3012 g_*^{-1/2} \frac{m_{Pl}}{T^2} \\
R \approx 3.699 \times 10^{-10} g_{*}(T)^{-1/3}\frac{\mathrm{MeV}}{T} \,.
\end{gather} 
Using these relations, we may compute 
\begin{align}
\frac{R_1 \ddot R}{\dot R^2} &= \frac{\left(\frac{t_1}{t}\right)^{1/2} \left(\frac{g(t)}{g\left(t_1\right)}\right)^{1/12} \left(13
   t^2 \dot g(t)^2-12 t g(t) \left(t \ddot g(t)+\dot g(t)\right)-36 g(t)^2\right)}{\left(t
   \dot g(t)-6 g(t)\right)^2} \\
      &= -\frac{1}{\hat \eta} \left[ \frac{-13 t^2 \dot g(t)^2+12 t g(t) \left(t \ddot g(t)+\dot g(t)\right)+36 g(t)^2}{\left(t
   \dot g(t)-6 g(t)\right)^2}\right] \\
   &= - {f_1(\hat \eta) \over \hat \eta} \,.
\end{align}
Above, we have defined 
\begin{equation}
f_1(\hat \eta) =  \frac{-13 t^2 \dot g(t)^2+12 t g(t) \left(t \ddot g(t)+\dot g(t)\right)+36 g(t)^2}{\left(t\dot g(t)-6 g(t)\right)^2} \,,
\end{equation}
where the right hand side is evaluated at the time $t$ corresponding to the conformal time $\hat \eta$. Similarly, we evaluate \begin{equation}
\frac{R_1^2 \dot R_1^2}{R^2 \dot R^2} = f_2(\hat \eta), \qquad \frac{\dot R_1^2}{\dot R^2} = \hat \eta^2 f_2(\hat \eta) \,,
\end{equation}
for
\begin{equation}
f_2(\hat \eta) = \frac{\left(\frac{g(t)}{g\left(t_1\right)}\right){}^{7/3} \left(t_1
   \dot g\left(t_1\right)-6 g\left(t_1\right)\right){}^2}{\left(t \dot g(t)-6
   g(t)\right)^2} \,.
\end{equation}
Finally, we define $f_3(\hat \eta) = m_a(\hat \eta(T))^2 / H_1^2 $. Combining these results, the equations of motion take the form 
\begin{gather}
 \phi_1'' + \left(\frac{3}{\hat \eta} - \frac{f_1(\hat \eta)}{\hat \eta}\right)\phi_1' - f_2(\hat \eta) \nabla^2 \phi_1 +\eta^2 f_2(\hat \eta) \left[ \tilde \lambda  \phi _1  \left(\phi _1^2+\phi _2^2 - 1\right) -\frac{f_3(\hat \eta) \phi _2^2 }{\left(\phi _1^2+\phi _2^2\right){}^{3/2}} \right] = 0 \\ 
\phi_2'' + \left(\frac{3}{\hat \eta} - \frac{f_1(\hat \eta)}{\hat \eta}\right)\phi_2' - f_2(\hat \eta) \nabla^2 \phi_2 +\eta^2 f_2(\hat \eta) \left[\tilde \lambda  \phi _2 \left(\phi _1^2+\phi _2^2 - 1\right) + \frac{f_3(\hat \eta) \phi _1 \phi _2}{\left(\phi _1^2+\phi _2^2\right)^{3/2}}\right]= 0.
\end{gather}
In the single-field formalism, these results are analogously applied to obtain
\begin{equation}
 \theta'' + \left(\frac{3}{\hat \eta} - \frac{f_1(\hat \eta)}{\hat \eta}\right)\theta' - f_2(\hat \eta) \bar \nabla^2 \theta +\eta^2 f_2(\hat \eta)f_3(\hat \eta) \sin \theta= 0 \,.
\end{equation}

\begin{figure*}[htb]
\includegraphics[width=.66\textwidth]{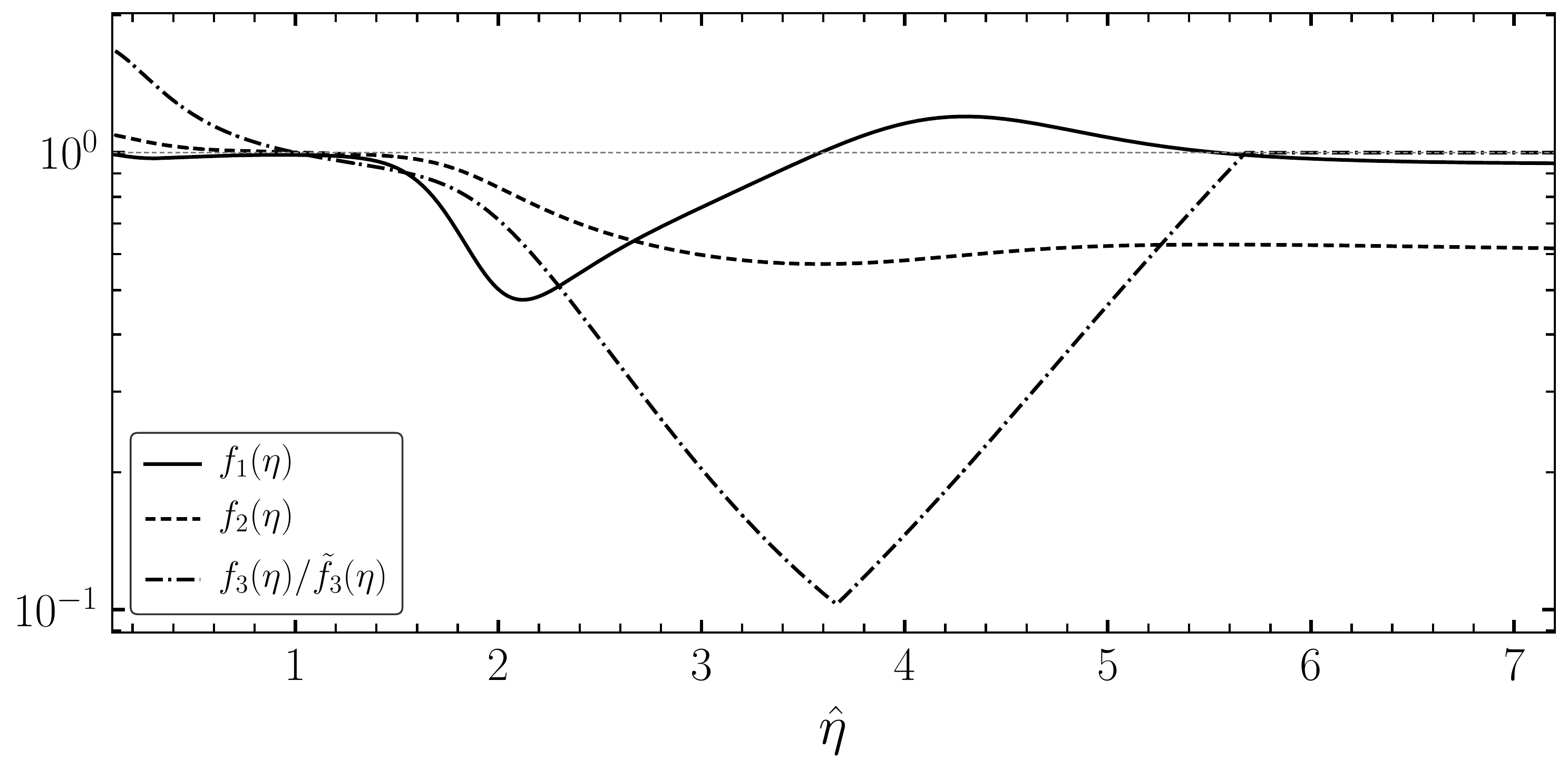}
  \caption{We depict the variation of $f_1$, $f_2$, and $f_3$ as a function of $\hat \eta$ over the relevant range of $\hat \eta$ for our simulations accounting for a varying $g_*$. For fixed $g_*$, we would expect $f_1$ and $f_2$ to be constant at value 1. We additionally show the behavior of $f_3$, which describes the evolution of the quantity $m_a(\eta)^2 / H_1^2$, normalized to $\tilde f_3$, wherein we compute $m_a(\eta)^2 / H_1^2$ assuming a fixed $g_*$. Assuming a fixed $g_*$ causes the axion to reach its zero-temperature value earlier in $\hat \eta$, but the ratio ultimately reaches unity as the same zero-temperature mass is reached.}
  \vspace{0.3cm}
  \label{fig:geff}
\end{figure*}

In Fig.~\ref{fig:geff} we show the functions $f_1$, $f_2$, and $f_3$ entering into the equations of motion as functions of $\hat \eta$.  Note that for $f_3$, we normalize against $\tilde f_3(\hat \eta)$, which we define to be $f_3$ but with a fixed $g_*$.  In the absence of a varying $g_*$, all of the curves appearing in Fig.~\ref{fig:geff} would be identically one.

To test the impact of an evolving $g_*$, we adopt the parametrization of $g_*$ from \cite{Wantz:2009it} and simulate, in two spatial dimensions, for an axion with $f_a = 4.83 \times 10^{15} \, \mathrm{GeV}$. When we assumed $g_*$ was constant, this axion reached its zero-temperature mass at $\hat \eta = 3.6$, but accounting for the changing value of $g_*$, the axion now reaches its zero-temperature mass at $\hat \eta \approx 5.5$. As before, we conclude our simulation at $\hat \eta = 7$, and we calculate a relic abundance that is $7.7\%$ smaller than it is in the fixed $g_*$ case. We note that this scenario represents something of a worst-case scenario for the impact of $g_*$ on the dynamics because $g_*$ varies significantly during the epoch where axion number density is not conserved for this choice of $f_a$, and so we adopt this as a quantification of the systematic error associated with adopting a fixed $g_*$.  For our target $f_a$, $g_*$ varies less, relative to the example illustrated, when the axion is in the non-linear regime and so we expect the effect of varying $g_*$ to be less important in this case. 

\section{Testing the Impact of the Mass Parametrization}
Precise details regarding the temperature dependence of the axion mass remain uncertain.
While we have chosen to use the parametrization of \cite{Wantz:2009it} with index $n = 6.68$ as done in \cite{Hiramatsu:2010yu, Kawasaki:2014sqa}, an alternative result is provided in \cite{Borsanyi:2016ksw}. In that work, an index of $n \approx 8.2$ is found at high temperatures, though 
we do note that an increasingly shallow dependence on $T$ is realized at lower temperatures. Motivated by power-law fits to this numerical result and informed by considerations of the changing number of degrees of freedom, recent works have taken an index of $n = 7.6$ in \cite{Klaer:2017ond} and $n = 7.3$ in \cite{Vaquero:2018tib} to study the axion field.  In this section, we use the extreme value $n = 8.2$ to estimate the maximal effect that uncertainties in the mass growth may have on the determination of the DM density.

We perform simulations for different $n$ in two spatial dimensions.  This is done for computational efficiency, and while we do not expect such a simplification to significantly affect the main conclusions we caution some care should be taken when interpreting these results for this reason.
We fix $\tilde \lambda = 5504$ and $f_a \approx  4.8 \times 10^{14} \, \mathrm{GeV}$.  This choice of $f_a$ corresponds to $\hat \eta_c = 3.6$ in the $n=6.68$ parametrization.  However, the value of $\hat \eta_c$ depends on our choice of $n$ and is $\hat \eta_c \approx 3.1$ for the choice of $n = 8.2$, since the mass grows faster in that case. We re-simulate with this alternative choice of index until $\hat \eta = 7$ and then recompute the present-day axion abundance by analytically transferring the simulation fields to the same late time physical temperature. We find that there is a $\sim$$10\%$ enhancement in the expected relic abundance with $n = 8.2$ versus $n = 6.68$.  This is somewhat surprising, considering that the analytic estimate predicts that higher $n$ should result in a lower DM abundance at fixed $f_a$.  To understand how this result affects the final determination of the axion mass, we fit the predicted scaling $\Omega_a \sim f_a^{(n+6) / (n+4)}$ for the DM abundance using $n = 8.2$ and find the $m_a$ that gives the correct DM abundance.
 The result is that with $n = 8.2$ we find that the $m_a$ that gives the correct DM abundance is enhanced by $\sim$27\% compared to the $n = 6.68$ case.  We account for this 27\% uncertainty as an additional systematic uncertainty in our final determination of the axion mass.

\section{Testing Deviations from the String Scaling Regime}

While our simulation was performed in two stages, it can be understood as a single simulation in which the PQ phase transition and the beginning of the QCD phase transition are separated by approximately an order of magnitude in temperature. By comparison, for a physically motivated hierarchy, we would expect these two epochs in our simulation to be separated by at least 11 orders of magnitude in temperature. As a result, our simulations might be expected to be highly unphysical. However, it has been conjectured that the axion field and associated defect network enters a scaling regime some time after the PQ phase transition (see, {\it e.g.},~\cite{Hiramatsu:2010yu}). If this conjecture is true and our field configuration has entered the scaling regime before the axion begins to oscillate, our simulation should be expected to give a good description of the physics of interest despite the abbreviated hierarchy. 

Recent work has found evidence for logarithmic deviations to the number of strings per Hubble patch in the scaling regime~\cite{Gorghetto:2018myk}.  In this section we confirm that we also observe such deviations.  This implies that we are not fully justified in taking the final state of our PQ-epoch simulation, fast-forwarding through the rest of the radiation dominated epoch to the QCD phase transition, and then restarting our simulation directly before the QCD phase transition.  This is because the axion-string network should change logarithmically during the evolution between the phase transitions.  Below, we provide evidence for the logarithmic deviation to scaling and then perform simulations to address the impact of this deviation on our determination of the axion mass $m_a$ and the spectrum of DM minihalos.

The average number of strings per Hubble patch is commonly defined by~\cite{Hiramatsu:2010yu,Fleury:2015aca,Gorghetto:2018myk}
\begin{equation}\label{eq:LogForm}
\xi(\tilde \eta)  = \frac{l(\tilde \eta) t(\tilde \eta)^2}{L(\tilde \eta)^3} \,,
\end{equation} 
where $l(\tilde \eta)$, $L(\tilde \eta)$, and $t(\tilde \eta)$ are the physical total string length in the box, the physical box length, and the physical time, each a function of $\tilde \eta$, respectively. 
We measure the string length by first identifying grid sites that are next to a string. This is achieved by forming a loop in each of the three dimensions around a test grid site. The grid site is flagged once at least one change larger than $\pi$ in the axion field between consecutive grid sites is found.
In a 2D slice this implies the 4 closest grid sites that surround the string core are tagged, such that we use the number of tagged grid sites divided by 4 as
a measure for the string length. Note that this is a rough estimate for the string length and more sophisticated methods exist~\cite{Hiramatsu:2010yu}.

\begin{figure*}[htb]
\includegraphics[width=.56\textwidth]{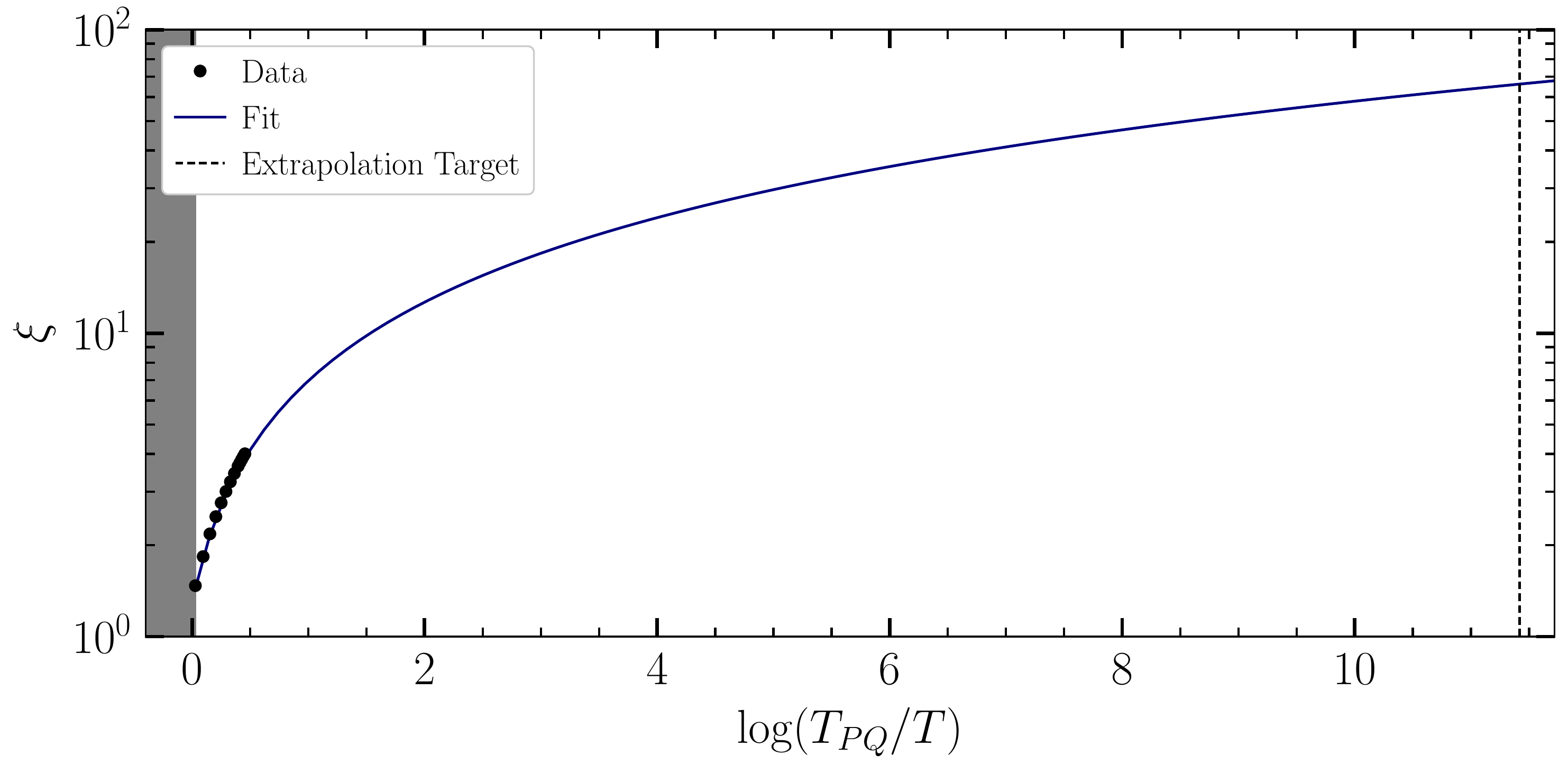}
  \caption{The string length parameter $\xi$ shown as a function of the ratio of simulation temperature $T$ to the temperature $T_{PQ}$ at which the PQ phase transition occurred, including the results of our fit to the functional form of~\eqref{eq:LogForm}. We observe significant evidence for logarithmic deviation from the scaling regime.  Extrapolating this result to the QCD phase transition (vertical dashed line) gives the prediction that $\xi$ should be around a factor of 15 higher at the beginning of the QCD phase transition than in the final state of our most realistic simulation.}
  \vspace{0.3cm}
  \label{fig:TotalStringLength}
\end{figure*}

\begin{figure*}[htb]
\includegraphics[width=.56\textwidth]{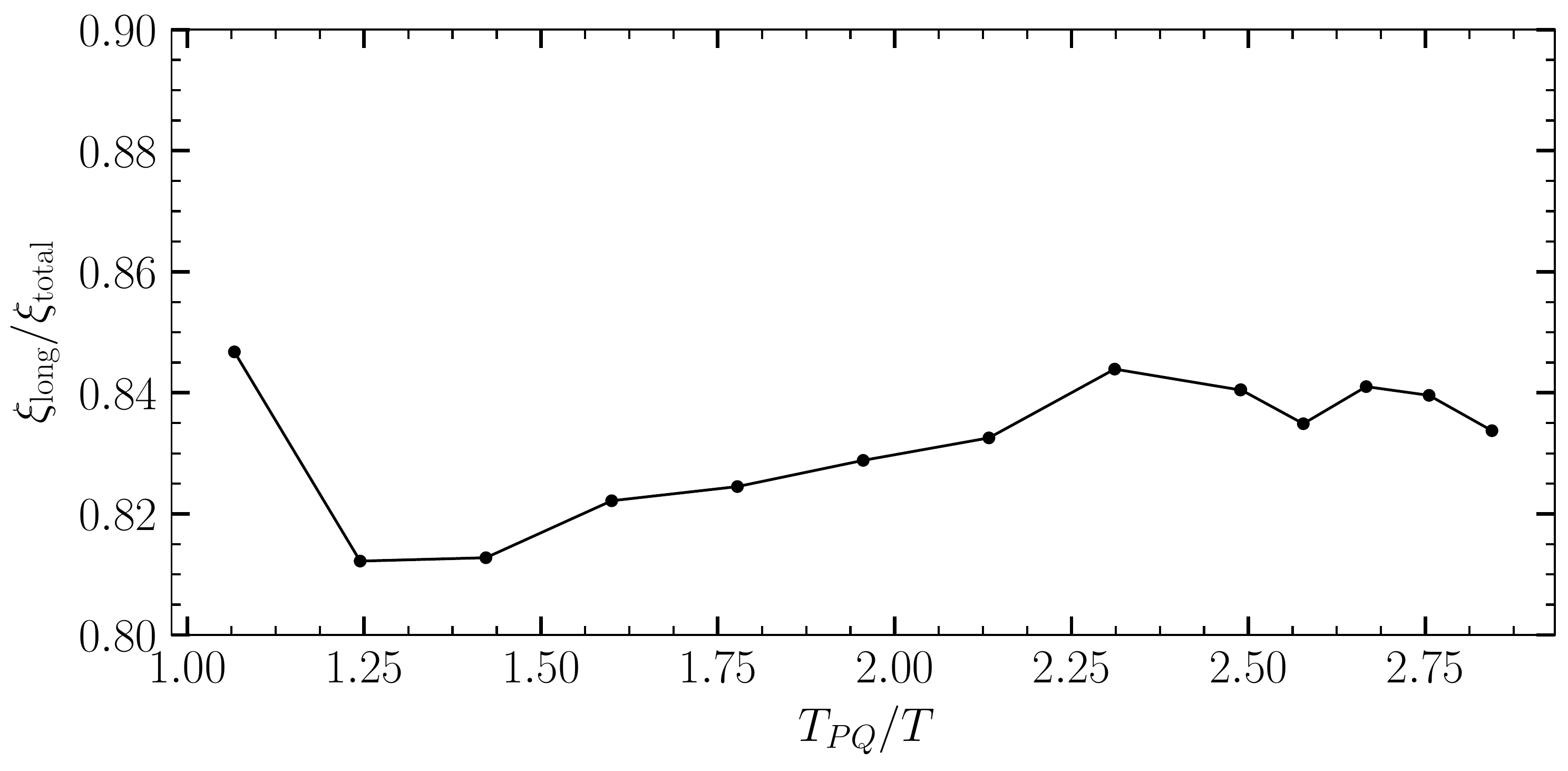}
  \caption{The fraction of the string length in super-horizon length strings. Like \cite{Gorghetto:2018myk}, we find roughly $80\%$ of the string length resides within long strings.}
  \vspace{0.3cm}
  \label{fig:LongStringLength}
\end{figure*}

We compute $\xi(\tilde \eta)$ at 13 points in $\tilde \eta$ in our PQ simulation, with results illustrated in Fig.~\ref{fig:TotalStringLength}.   
As in~\cite{Gorghetto:2018myk}, we find that $\xi$ depends logarithmically on $\tilde \eta$ after the PQ phase transition.  Note that the shaded region denotes $\tilde \eta$ before the PQ phase transition, where it does not make sense to talk about axion strings.  We fit the model
\begin{equation}
\xi = \alpha \log\left(\frac{T}{T_{PQ}}\right) + \beta
\end{equation}
to the $\{ \tilde \eta, \xi \}$ data, where $T_{PQ}$ denotes the temperature of the PQ phase transition, and we
find $\alpha \approx -2.60$ and $\beta \approx 1.27$.  Note that our values for $\xi$ at comparable $\tilde \eta$ are significantly larger than those found in \cite{Gorghetto:2018myk}.  Part of this discrepancy could be due to the way in which we measure string length versus in that work, which may introduce an overall rescaling between our two results.
We are in good agreement, however, with \cite{Gorghetto:2018myk} regarding the distribution of string length in long and short strings.  As in that work, we find that approximately $80\%$ of the string length resides in long strings, much larger than a Hubble length, at all times in our simulation as seen in Fig.~\ref{fig:LongStringLength}.

Since we do observe logarithmic scaling violations, it is important to determine the impact of these corrections on the minihalo mass spectrum and the DM relic abundance.  In particular, we find that $\xi$ should be around a factor of 15 higher at the QCD phase transition than it is for the final state of our most realistic PQ-epoch simulation.
The string density $\xi$ at the beginning of the QCD simulation depends on the simulation box size as $\xi_{\rm QCD} \propto L_{\rm QCD}^{-2}$, where $L_{\rm QCD}$ is the box size in units of $1/(a_1 H_1)$ when $H = m_a$, for a fixed initial state.  Thus by performing new simulations with the same initial conditions and run parameters as our fiducial analysis (namely $\hat \eta_c = 3.6, \tilde \lambda = 5504$, and starting at $\hat \eta_i = 0.4$) but changing $L_{\rm QCD}$ from 4 to $L_{\rm QCD} = 3$ and $L_{\rm QCD}=2$, we may enhance $\xi_{\rm QCD}$ by a factor of $16/9$ and $4$, respectively, compared to our fiducial simulation.  While these value still fall short of the physically motivated enhancement $\sim$$15$, such simulations still allow us to see if there is a trend in how $\xi$ affects observables such as the DM density. We do caution that modifying $L_{\rm QCD}$ in this way is somewhat unphysical as it changes the horizon entry status of modes in the simulation box from the end of the PQ simulation as compared to the beginning of the QCD simulation. 

 \begin{figure*}[htb]
\includegraphics[width=.66\textwidth]{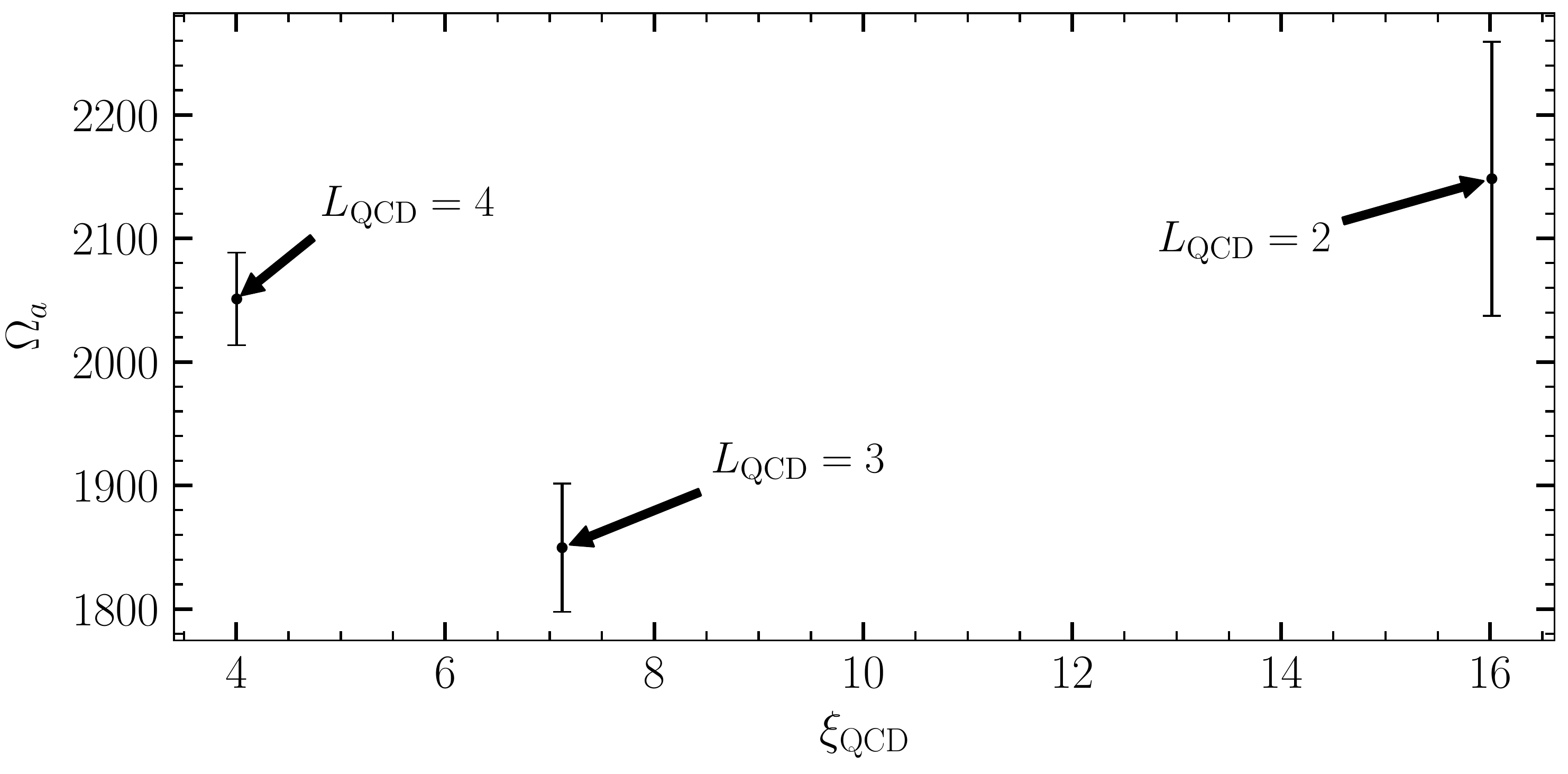}
  \caption{The present day axion abundance as a function of the string density parameter $\xi$ at the beginning of the QCD simulation at $\hat \eta_i = 0.4$. Individual data points are labeled by their box length $L_{\rm QCD}$.  The error bars are estimates of the statistical uncertainties, and no clear trend is visible in the data.}
  \vspace{0.3cm}
  \label{fig:StringDensityRelicAbundance}
\end{figure*}

 \begin{figure*}[htb]
\includegraphics[width=.56\textwidth]{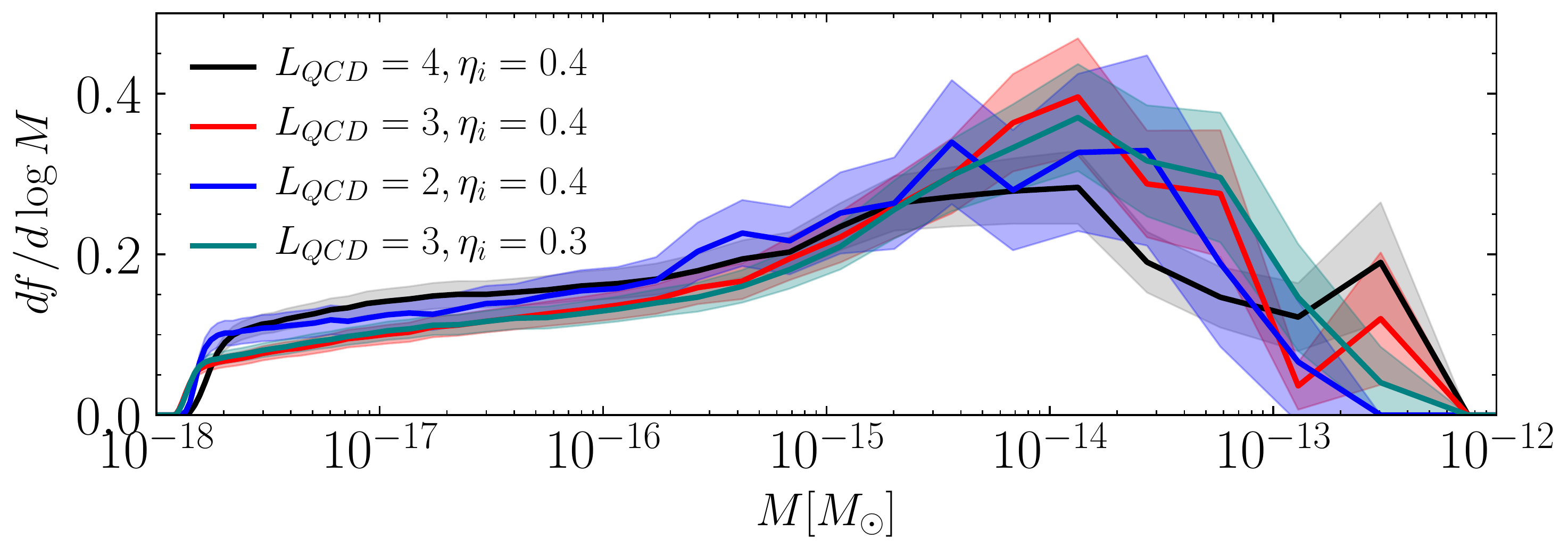}
  \caption{Differential mass spectrum as a function of the minihalo mass $M$ for different box sizes. Error bands include statistical errors and the uncertainty on the overall normalisation.}
  \vspace{0.3cm}
  \label{fig:ReinterpSims}
\end{figure*}

The results of varying $L_{\rm QCD}$ in order to modify $\xi$ are shown in Fig.~\ref{fig:StringDensityRelicAbundance}, where we see no discernible trends in the dependence of the relic abundance on $\xi$.  Note that the uncertainties in Fig.~\ref{fig:StringDensityRelicAbundance} are estimates of the statistical uncertainty.  As the box size gets smaller the statistical uncertainty increases.  However, we caution that these are estimates only, as we have not run multiple independent simulations for each box size due to computational limitations.  It is possible that the true uncertainties at small box sizes are larger than indicated due to the fact that there are a small number of {\it e.g.} domain walls that form these cases.  Still, to be maximally conservative given the available datasets, we estimate the difference between the $L_{\rm QCD} = 2$ and $L_{\rm QCD} = 3$ values for $\Omega_a$ as a systematic uncertainty induced from the deviation to scaling.  However, we cannot be sure that this difference is not a result of statistics or from the way in which we artificially mock up initial conditions with higher $\xi$ values.  The systematic uncertainty we assign from these tests is $15\%$ correlated between different $f_a$ points.   

We show in Fig.~\ref{fig:ReinterpSims} the impact on the minihalo mass spectrum. Again, all simulations largely agree within their error bands (estimated from statistical uncertainties), indicating that an increase in $\xi$ has only a marginal effect on the late-time axion field.   
Note that computational resources limit us to just these three additional simulations, and we leave a more detailed investigation of the dependence of $\Omega_a$ on $\xi$ to future work.

\end{document}